\documentclass[%
superscriptaddress,
 preprint,
 amsmath,amssymb,
 aps,
 jcp,
 letterpaper,
]{revtex4-2}

\usepackage{graphicx}% Include figure files
\usepackage{dcolumn}% Align table columns on decimal point
\usepackage{bm}% bold math
\newcolumntype{.}{D{.}{.}{8}}

\usepackage[utf8]{inputenc}
\usepackage{physics}
\usepackage{amsmath, amssymb}
\usepackage{tabularx,booktabs,array,dcolumn}
\usepackage{setspace}
\usepackage{vmargin}
\usepackage{xcolor}
\usepackage{graphicx,psfrag,subfigure}

\usepackage{float}
\usepackage[all]{xy}
\usepackage{multirow}

\usepackage{bm}
\usepackage{mathtools}
\usepackage{physics}

\newcommand{\bos}[1]{\boldsymbol{#1}}

\newcommand{\cm}{cm$^{-1}$}

\newcommand{\som}{SOM}

\usepackage[unicode]{hyperref}
\hypersetup{
   unicode=true,          % non-Latin characters in Acrobat??s bookmarks
   plainpages=false,
   colorlinks=true,       % false: boxed links; true: colored links
   citecolor=blue,        % color of links to bibliography
}

\begin{document}

\title{%
  Performance of a black-box-type rovibrational method in comparison 
  with a tailor-made approach: case study for the methane-water dimer
}

\author{Alberto Mart\'in Santa Dar\'ia}
\author{Gustavo Avila}
\author{Edit M\'atyus}
\email{edit.matyus@ttk.elte.hu}

\affiliation{Institute of Chemistry, 
ELTE E\"otv\"os Lor\'and University
P\'azm\'any P\'eter s\'et\'any 1/A
1117 Budapest, Hungary}

\date{\today}
\begin{abstract}
  \noindent 
  The present work intends to join and respond to the excellent and thoroughly documented rovibrational study of 
  [X.-G. Wang and T. Carrington, Jr., J. Chem. Phys. 154, 124112 (2021)]
  that used an approach tailored for floppy dimers with an analytic dimer Hamiltonian and 
  a non-product basis set including Wigner $D$ functions.
  It is shown in the present work that the GENIUSH black-box-type rovibrational method can approach the performance of the tailor-made computation for the example of the floppy methane-water dimer. 
  Rovibrational transition energies and intensities are obtained in the black-box-type computation with `only' 2-3 times larger basis set sizes and in an excellent numerical agreement in comparison with the more efficient tailor-made approach.
\end{abstract}

\maketitle

%%%%%%%%%%%%%%%%%%%%%%%%%%%%%%%%%%%%%%%%%%%%%%%%%%%%%%%%%%%%%%%%%%%%%%%%%%%%%%%%%%%%%%%%%%%%%
%
% Introduction
%
%%%%%%%%%%%%%%%%%%%%%%%%%%%%%%%%%%%%%%%%%%%%%%%%%%%%%%%%%%%%%%%%%%%%%%%%%%%%%%%%%%%%%%%%%%%%%
\clearpage
\section{Introduction}
\noindent
Electronic structure theory has general program packages that work for $n$ electrons, where $n$ is defined by the user together with the configuration of the clamped nuclei.

Will computational (ro)vibrational spectroscopy methodology reach this level of automatization? Does it need it to reach this and why? 
One may argue, that careful gas-phase spectroscopy measurements conducted in order to explore the structure and internal dynamics of molecular systems typically assume months or even years of systematic experimental work (development of sample preparation, source and detector architectures, uncertainty estimation, error control, etc.).
It is not a typically high-throughput field studying one molecule after the other at an extraordinary pace. 
At the same time, the experiments may produce large sets of precise data that provide a highly detailed characterization
of the quantum dynamical features of the system.

Regarding computational molecular rovibrational spectroscopy, it is a natural aim for a mathematically formulated theory to have an approach, an algorithm, and a computer program
that is generally applicable, limited `only' by the available computational resources. 
For solving the rovibrational Schrödinger equation, it should be an general $N$-particle approach, where $N$ is the number of atomic nuclei (although,
for the moment only finite many $N$ possible values are computationally feasible). In this field, it is another necessary condition to allow the user
to specify the internal coordinates and the body-fixed frame best suited for the computations. 

Development of Ref.~\cite{MaCzCs09}, following earlier work in the field \cite{MeGu69,Lu00,Lu03,LaNa02,YuThJe07}, 
was led by this idea in spite of the fact that there were already excellent approaches available specifically designed for special types of molecular systems (with a given number of nuclei and specific coordinates), the list includes, for example Refs.~\cite{BrAvSuTe83,BaLi89,SuTe91,BrCa94,Le94,AlCl94,ZhWuZhDiBa95,Ml02,WaCa04}. The numerical advantage of an $N$-particle method was not at all obvious over the performance of tailor-made approaches. 
For semi-rigid systems, the Eckart--Watson Hamiltonian \cite{Wa68} was available as a general $N$-particle Hamiltonian, and for which efficient solution techniques have been developed using basis pruning \cite{WhHa75,MM2}. 
If basis pruning can be efficiently realized, then the grid pruning techniques can be employed \cite{tc-gab1,tc-gab2}
to milden the curse of dimensionality. Most recently, basis and grid pruning techniques have been used to describe efficiently the semi-rigid part of floppy complexes in full dimensionality \cite{AvMa19,AvMa19b,AvPaCzMa20}.

Floppy molecular systems with multiple large-amplitude motions was and (is still) an open challenge for the field, so it was natural to ask the following questions about a novel black-box-type approach.
1) Can the black-box-type rovibrational approach tackle floppy systems? 
2) Can we come close in efficiency to tailor-made approaches for floppy systems? 
Although singular regions can be explored numerically, there are difficulties connected to the selection and
use of good basis functions and integration grids
without an analytic knowledge of the kinetic energy operator (KEO) and its matrix elements \cite{AvMa19}. 
3) Is it necessary to group the coordinates in a certain fashion and use specifically coupled basis functions, 
or is it possible to continue using the simpler direct product basis and grid representations for a start, that can be pruned perhaps at a next stage of the development? 

In the present work, we will show for the example of the methane-water dimer that the answer to all three questions, 1--3), is `yes'.

We use the methane-water dimer as an example system, because Wang and Carrington \cite{WaCa21} recently reported and carefully documented their work using a tailor-made dimer Hamiltonian, Wigner $D$ basis functions, and the symmetry adapted Lanczos eigensolver for computing rovibrational transitions of this dimer. They have carefully compared their computational results with earlier black-box-type rovibrational computations carried out for this system \cite{SaCsAlWaMa16,SaCsMa17,dimers}, and pointed out that even an excessively large direct-product basis and grid provided not sufficiently converged rovibrational energies, with inaccuracies on the order of 0.05--1.5~\cm. Since the publication of the first applications of the GENIUSH program for molecular dimers, we have gathered more experience in treating floppy complexes, and using this development and experience we wish to complement this comparison.

The methane-water dimer is also a chemically and spectroscopically important example. It is the simplest model for water-hydrocarbon interactions, there are high-resolution far-infrared \cite{DoSa94} and microwave spectroscopic data \cite{SuFrLoKa94} available for the system, while the analysis of the congested spectrum was difficult and only partially possible by effective Hamiltonians.

In this work, we will show that using an order of magnitude \emph{smaller} but better basis and grid representation, than in earlier work \cite{SaCsAlWaMa16,SaCsMa17,dimers}, we can converge the rovibrational energies to 10$^{-3}$~\cm\ with a basis and grid that is `only' 2-3 times larger, than the coupled basis used by Wang and Carrington \cite{WaCa21}. 
The tailor-made approach remains to be more efficient, but not by orders of magnitude. The efficiency of the GENIUSH computation can be further enhanced by exploiting part of the rich symmetry features \cite{FaQuCs17} of this particular complex.
We think that the level of convergence achieved in this work is much beyond the range of the typical approximations
underlying the current computational rovibrational spectroscopy framework (quality of the potential energy surface, Born--Oppenheimer and non-relativistic approximations).

Finally, we may ask whether we can expect any fundamental or numerical advantage from developing general rovibrational approaches, apart from fulfilling a mathematical `necessity'.

First of all, we may aim for automatically defining internal coordinates that are optimal or near optimal for a particular computation (system and energy range). 

Furthermore, for a floppy molecular system, a fundamental and numerically important open problem is finding the optimal body-fixed frame, 
or at least a body-fixed frame that is good enough or better than another one. For small-amplitude vibrations, the Eckart frame is known to be an excellent choice in minimizing the rotational and vibrational problem for low-energy rovibrational states, and thus, it allows us to use the $J=0$ vibrational eigenfunctions as a basis for $J>0$ computation \cite{NMD10} in the same energy range. For higher excited semi-rigid systems and especially for floppy systems, we do not have any practical approach for finding a good or at least a `better' frame, and it may be the numerical KEO approach used also in the GENIUSH program that will allow us to optimize the molecular frame `on-the-fly' over a grid \cite{AvMSDMe21}.

Having this perspective in mind, after a short theoretical introduction (Sec.~\ref{ch:theo}), we report rovibrational energies (Secs.~\ref{ch:enlev} and \ref{ch:comparison}), transitions and line strengths  (Secs.~\ref{ch:linestr}, \ref{ch:firspectrum}, and \ref{ch:mirspectrum}) for the example of methane-water obtained with the numerical KEO approach of GENIUSH and using the potential energy surface (PES) developed by Akin-Ojo and Szalewicz \cite{AOSz05}.

%%%%%%%%%%%%%%%%%%%%%%%%%%%%%%%%%%%%%%%%%%%%%%%%%%%%%%%%%%%%%%%%%%%%%%%%%%%%%%%%%%%%%%%%%%%%%
%
% Theory
%
%%%%%%%%%%%%%%%%%%%%%%%%%%%%%%%%%%%%%%%%%%%%%%%%%%%%%%%%%%%%%%%%%%%%%%%%%%%%%%%%%%%%%%%%%%%%%
\clearpage
\section{Theoretical description and computational details \label{ch:theo}}
\noindent
The quantum dynamical computations were carried out by using the GENIUSH \cite{MaCzCs09,FaMaCs11} computer program. This program package has been used already to study a number of semi-rigid and floppy molecular systems \cite{FaCsCz13,14FaMaCs,SaCsAlWaMa16,ArNOp,FaQuCs17,SaCs16,SaCsMa17,dimers,FeMa19,AvMa19,AvMa19b,AvPaCzMa20,fad21}, so here we only shortly summarize the theoretical background.  The general rovibrational Hamiltonian  \cite{MeGu69,MeJMS79,Lu00,Lu03,LaNa02,YuThJe07}
  \begin{equation}
   \label{eq:hamil}
   \begin{split}
    & \hat{H} = \frac{1}{2} \sum_{k=1}^D \sum_{l=1}^D \tilde{\text{g}}^{-1/4} \hat{p}_k G_{kl}\tilde{\text{g}}^{1/2}  \hat{p}_l\tilde{\text{g}}^{-1/4} \\
    & + \frac{1}{2} \sum_{k=1}^D \sum_{a=1}^3(\tilde{\text{g}}^{-1/4} \hat{p}_k G_{k,D+a}\tilde{\text{g}}^{1/4} + \tilde{\text{g}}^{1/4} G_{k,D+a}\hat{p}_k \tilde{\text{g}}^{-1/4})\hat{J}_a \\
    & + \frac{1}{2} \sum_{a=1}^3 G_{D+a,D+a}\hat{J}_a^2\\
    & + \frac{1}{2} \sum_{a=1}^3 \sum_{b>a}^3 G_{D+a,D+b}[\hat{J}_a,\hat{J}_b]_+ + \hat{V}
    \end{split}
   \end{equation}
is implemented in the GENIUSH program. In Eq.~\ref{eq:hamil}, $\hat{J}_a$ ($a=1(x),2(y),3(z)$) are the body-fixed total angular momentum operators and $\hat{p}_k=-i\partial/ \partial q_k$ with the $q_k\ (k=1,2,\ldots,D)$ internal coordinates. The $G_{kl}=(\textbf{g}^{-1})_{kl}$ coefficients and $\tilde{\text{g}}=\text{det}(\textbf{g})$ are determined from the rovibrational $\bos{g}\in\mathbb{R}^{(D+3)\times(D+3)}$ matrix, defined as follows,
\begin{equation}
  g_{kl}= \sum_{i=1}^N m_i \textbf{t}^\text{T}_{ik} \textbf{t}_{il}; 
   \quad\quad\quad k,l = 1,2,...,D+3 
  \label{eq:gmxt}
\end{equation}
with
\begin{equation}
\textbf{t}_{ik} = \frac{\partial \textbf{r}_i}{\partial q_k}; 
   \quad\quad\quad k,l = 1,2,...,D
   \label{eq:tvib}
\end{equation}
\begin{equation}
\textbf{t}_{i,D+a} = \textbf{e}_a \times \textbf{r}_i; 
   \quad\quad\quad a = 1(x),2(y),3(z) \; ,
   \label{eq:trot}   
\end{equation}
where $\textbf{r}_i$ are the body-fixed Cartesian coordinates for the $i$-th atom and $\textbf{e}_a$ represent the body-fixed unit vectors. 

\begin{figure}
    \begin{center}
      \includegraphics[scale=0.6]{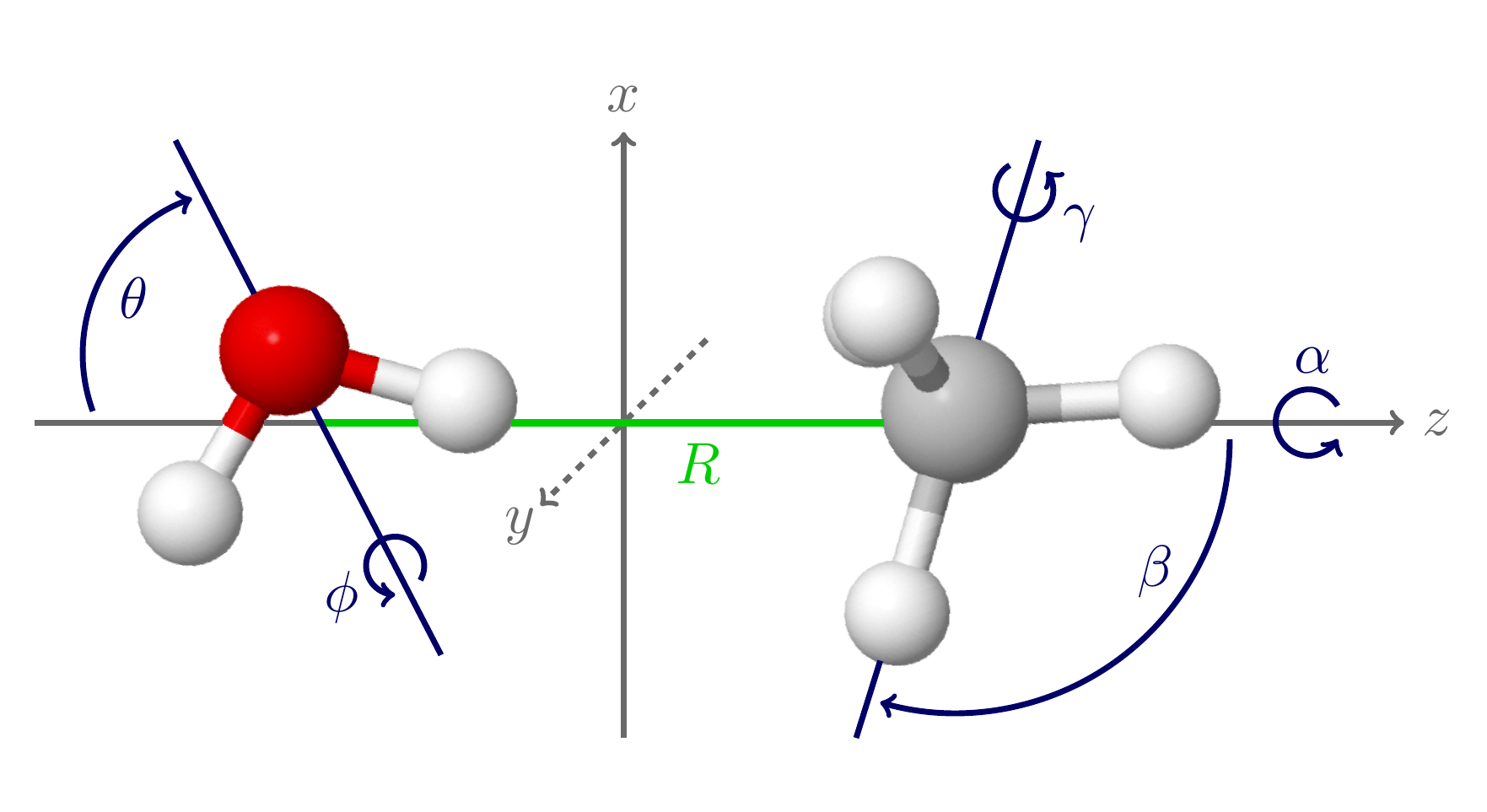}
    \end{center}      
    \caption{
      Definition of the intermolecular coordinates, $(R,\theta,\phi,\alpha,\beta,\gamma)$ of the CH$_4$--H$_2$O dimer.
      \label{fig:dimer}      
      }
\end{figure}
In the CH$_4$--H$_2$O dimer, the intermolecular degrees of freedom ($D = 6$, Fig.~\ref{fig:dimer})  are defined with the following internal coordinates: 
the $R \in [0,\infty)$ distance between the centers of mass of the monomers; 
two angles, $\cos{\theta} \in [-1,1]$ and $\phi \in [0,2\pi)$, to describe the orientation of H$_2$O; 
and 
three angles, $\alpha \in [0,2\pi)$, $\cos{\beta} \in [-1,1]$ and $\gamma \in [0,2\pi)$, to describe the orientation of CH$_4$.
The monomer structures are fixed at the effective vibrational structures used for the development of the PES in Ref.~\cite{AOSz05} and in earlier rovibrational computations \cite{SaCsAlWaMa16,WaCa21}. 
For completeness, we repeat here the values of the constrained  
$\lbrace \text{bond length, angle}\rbrace$ values,  that
are 
$\lbrace r(\text{O--H}) = 0.9716257\ \text{\r{A}}, \alpha(\text{H-O-H}) = 104.69^{\circ}\rbrace$ and 
$\lbrace r(\text{C-H}) = 1.099122\ \text{\r{A}}, \cos\alpha(\text{H-C-H}) = -1/3\ ^{\circ}\rbrace$, for the water and the methane fragments, respectively.
We used the same atomic masses as in Refs.~\cite{SaCsAlWaMa16} and \cite{WaCa21}: $m(\text{H}) = 1.007825\ \text{m}_\text{u}$, $m(\text{C}) = 12\ \text{m}_\text{u}$ and $m(\text{O}) = 15.994915\ \text{m}_\text{u}$. 
Since, we use the same PES, constrained coordinates, and nuclear (atomic) mass values as Ref.~\cite{WaCa21} (and also Ref.~\cite{SaCsAlWaMa16}) direct comparison of the results is possible.

\begin{figure}
    \begin{center}
      \includegraphics[scale=0.6]{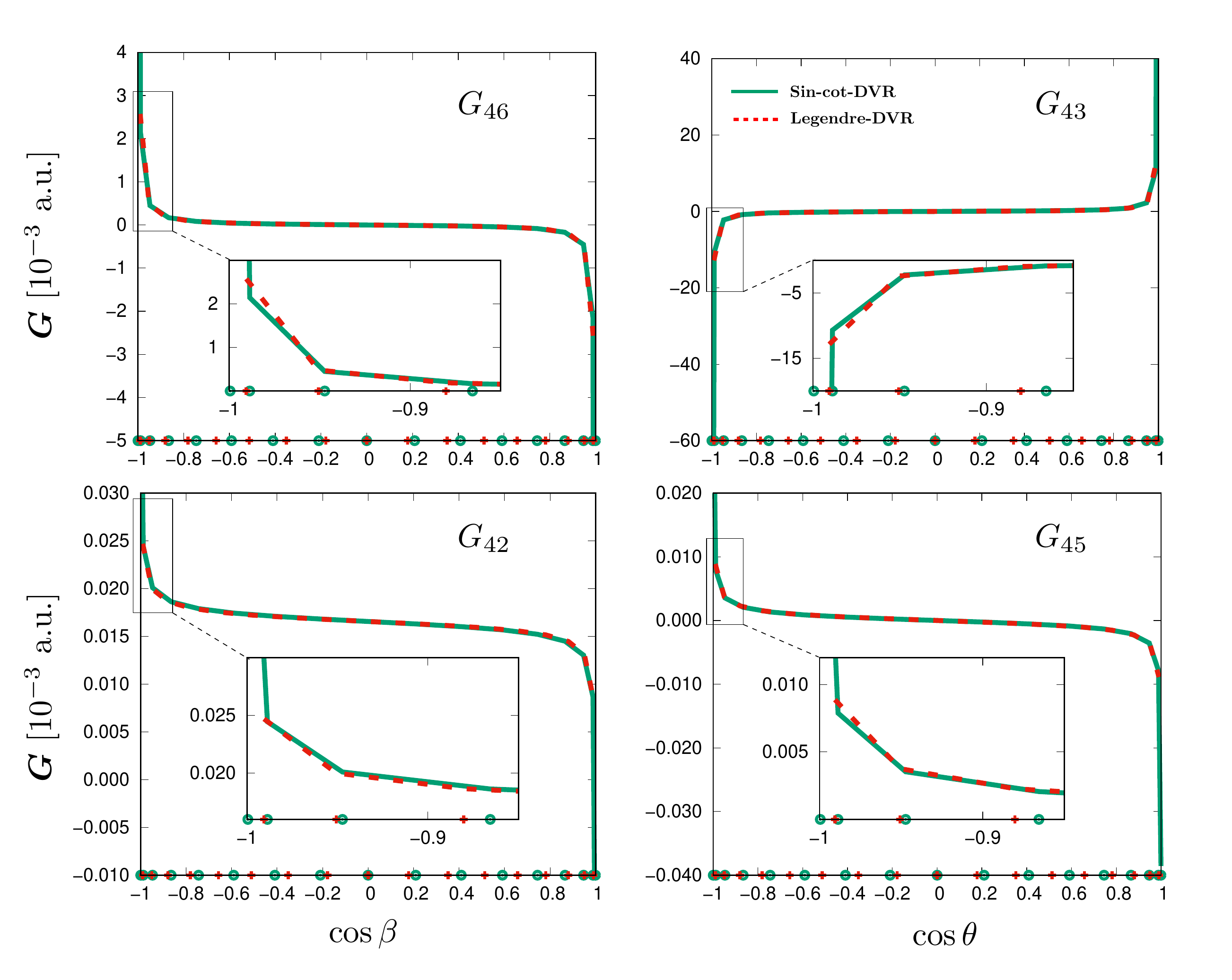}
    \end{center}      
    \caption{%
      Selected elements of the $\bos{G}$ matrix shown along the singular coordinates, $\cos\beta$ and $\cos\theta$ (while all other coordinates are fixed). The Legendre and sine-cot-DVR grid points (17 of each) are plotted on the $x$ axis of each subfigure. Both sets of points approach the $\pm 1$ singular points, but the sine-cot-DVR points have a higher density near the singularities. 
     \label{fig:Gmatrix}
      }
\end{figure}
In the present work, the matrix representation of the Hamiltonian is constructed using a direct product discrete variable representation (DVR) \cite{00LiCa} for the vibrational degrees of freedom. For our curvilinear coordinate representation (Fig.~\ref{fig:dimer}), the KEO has singularities at $\cos\theta = \pm1$ and $\cos\beta = \pm1$. 
Plain Legendre DVR can be used for these coordinates, but the convergence of the (ro)vibrational energies is slow and an excessive number of grid points is needed even for a moderate level of convergence \cite{SaCsAlWaMa16,SaCs16,dimers}. The representation is fundamentally correct, since we have repulsive singularities in the KEO {Figure~\ref{fig:Gmatrix}}, but it is computationally inefficient. (We note that while in Fig.~\ref{fig:Gmatrix} both positive and negative singular values can be seen, the integrals are always positive.)
This behaviour was correctly pointed out by Wang and Carrington in Ref.~\cite{WaCa21}. 
A direct product basis and grid with Legendre DVR used in the GENIUSH computations \cite{SaCsAlWaMa16,SaCs16,dimers}, which was 20 times larger than the coupled basis representation with Wigner $D$ functions of Wang and Carrigton \cite{WaCa21},
was sufficient to converge the rovibrational states in the ZPV splitting range only with a convergence error as large as 0.44~\cm\ and an erroneous split for some of the degenerate levels. At the same time, we note that this split did not hinder the unambiguous molecular symmetry group assignment of the GENIUSH results using the coupled-rotor decomposition scheme \cite{SaCsMa17}.

This convergence problem was intriguing for a full(12)-dimensional computation of the methane-argon complex studied by two of us \cite{AvMa19}. In order to make the 12D computation feasible, the more efficient, sine-cot-DVR developed by Schiffel and Manthe \cite{ScMa10} was used for the singular coordinate (that was the second Euler angle in that system). 
The sine-cot-DVR was developed by Schiffel and Manthe \cite{ScMa10} to have a
a more efficient representation of the quantum mechanical motion along the singular coordinate. 
(We note that the singularity is present only for $K\neq0$ in the $|J,K\rangle$ basis representation of Ref.~\cite{ScMa10}.)
It  uses a polynomial series of cosine, and optionally also sine, functions to build a DVR. In the present work, we obtained the best results when two sine functions were also included, and the `final' computational parameters are summarized in Table~\ref{tab:param}.

\begin{table}%[htbp]
%\scriptsize{
\caption{%
  Coordinate intervals and representations used in the GENIUSH rovibrational computations. 
}
  \label{tab:param}
\begin{tabular}{@{}l c@{\ \ \ } c@{\ \ } cc@{\ \ }c@{}}
\hline\\[-0.35cm] 
\hline\\[-0.35cm] 
  \raisebox{-0.20cm}{Coord.} & 
  \raisebox{-0.20cm}{GM$^\text{a}$} & 
  \multicolumn{3}{c}{DVR} & 
  \raisebox{-0.20cm}{No. points}  \\[-0.20cm]
\cline{3-5}\\[-0.35cm]
           &                   & type &  & interval &  \\
\hline\\[-0.35cm]
$R\ [\textup{\r{A}}]$ & 3.464 & PO-Laguerre$^\text{b}$&  & [2.5,6.0] & 15 \\
%$\cos\theta\ [^{\circ}]$  & 116.190 & Sin-Cot-DVR$^\text{c}$ &  & [1.0,179.0] & 17 \\
$\theta\ [^{\circ}]$  & 116.190 & Sine-Cot-DVR$^\text{c}$ &  & (0,180) & 17 \\
$\phi\ [^{\circ}]$   & 90.000 & Fourier &  & [0,360] & 15 \\
$\alpha\ [^{\circ}]$ & 297.460 & Fourier &  & [0,360] & 9 \\
%$\beta\ [^{\circ}] $   & 113.050 & Sin-Cot-DVR$^\text{c}$ &  & [1.0,179.0] & 17 \\
$\beta\ [^{\circ}] $   & 113.050 & Sine-Cot-DVR$^\text{c}$ &  & (0,180) & 17 \\
$\gamma\ [^{\circ}]$ & 293.010 & Fourier &  & [0,360] & 23 \\
\hline\\[-0.4cm] 
\hline
\end{tabular}
\begin{flushleft}
$^\text{a}$~Internal coordinates values for the global minimum (GM) structure of the AOSz05 PES \cite{AOSz05}. \\
$^\text{b}$~Potential-optimized DVR using 300 points.\\
$^\text{c}$~$\cos\beta$ and $\cos\theta$ were the active coordinates and the sine-cot-DVR was constructed with two sine functions.\\
\end{flushleft}
%}
\end{table}

%%%%%%%%%%%%%%%%%%%%%%%%%%%%%%%%%%%%%%%%%%%%%%%%%%%%%%%%%%%%%%%%%%%%%%%%%%%%%%%%%%%%%%%%%%%%%
%
% Energy levels
%
%%%%%%%%%%%%%%%%%%%%%%%%%%%%%%%%%%%%%%%%%%%%%%%%%%%%%%%%%%%%%%%%%%%%%%%%%%%%%%%%%%%%%%%%%%%%%
\clearpage
\section{Energy levels and convergence tests \label{ch:enlev}}
\noindent
First, the convergence of the results has been tested with respect to the number of grid points for every coordinate using a series of reduced-dimensional ($D<6$) and full-dimensional ($D=6$) computations. Table~\ref{tab:gridsize} highlights the $10^{-3}$~\cm\ convergence of the first 20 vibrational states for the selected grid parameters (Table~\ref{tab:param}). 

In Table~\ref{tab:gridsize}, we can observe a small, $10^{-3}$~\cm\ artificial split of some of the triply degenerate levels, which is due to the fact that the underlying grid and basis do not respect the full symmetry of the dimer, hence certain symmetry features are converged numerically by enlargement of the grid.
At the same time, looking at this level of convergence, we felt important to check three additional computational parameters that are listed in the following paragraphs and summarized in Table~\ref{tab:tests}.

\begin{table}[htbp]
\caption{%
  Convergence test of the vibrational ($J=0$) energy levels, in \cm, referenced to the zero-point vibrational energy in the methane-water dimer computed with the GENIUSH program and the AOSz05 PES \cite{AOSz05}. 
\label{tab:gridsize}  
}
\begin{tabular}{@{}r@{\ \ } r@{\ \ }r@{\ \ } r@{}}
\hline\\[-0.35cm]
\hline\\[-0.35cm]
J0.$n$ & 
\multicolumn{1}{c}{$\tilde{\nu}$ $^\text{a}$}	&
\multicolumn{1}{c}{$\delta^{(-2)}$ $^\text{b}$} &
\multicolumn{1}{c}{$\delta^{(+2)}$ $^\text{c}$} \\
\multicolumn{1}{l}{$N_\text{b}:$}	&
\multicolumn{1}{c}{$1.4\cdot 10^7$}	&
\multicolumn{1}{c}{$5.6\cdot 10^6$} &
\multicolumn{1}{c}{$5.9\cdot 10^7$} \\
\hline\\[-0.35cm]
1	&	206.81	&	0.001	&	0	\\
2	&	4.764	&	0.001	&	0	\\
3	&	4.764	&	0.001	&	0	\\
4	&	4.765	&	$-$0.001	&	0.001	\\
5	&	6.992	&	0.002	&	0	\\
6	&	11.25	&	0.002	&	0	\\
7	&	11.251	&	0.003	&	0.001	\\
8	&	11.251	&	0	&	0	\\
9	&	29.033	&	$-$0.001	&	0	\\
10	&	29.034	&	$-$0.001	&	0.001	\\
11	&	29.034	&	$-$0.001	&	0	\\
12	&	32.636	&	0.003	&	0	\\
13	&	32.637	&	0	&	0	\\
14	&	32.637	&	0	&	0	\\
15	&	32.711	&	0.002	&	0	\\
16	&	32.712	&	0.001	&	0.001	\\
17	&	32.712	&	0.001	&	0	\\
18	&	34.41	&	0	&	0	\\
19	&	35.92	&	0	&	0	\\
20	&	35.92	&	0	&	0	\\
\hline\\[-0.35cm]
\hline\\[-0.35cm]
\end{tabular}
\begin{flushleft}
$^\text{a}$~Obtained with the (15,17,15,9,17,23) grid documented in Table~\ref{tab:gridsize}. \\
$^\text{b}$~$\delta^{(-2)}=\tilde\nu-\tilde\nu^{(-2)}$, where $\tilde\nu^{(-2)}$ was obtained with the same type of grid as in (a), but with fewer points: (13,15,13,9,15,21). \\
$^\text{c}$~$\delta^{(+2)}=\tilde\nu-\tilde\nu^{(+2)}$, where $\tilde\nu^{(+2)}$ was obtained with the same type of grid as in (a), but with more points: (17,19,17,11,19,25). \\
\end{flushleft}
\end{table}

\paragraph{Testing the finite differences calculation of the vibrational t-vectors}
We have tested the accuracy of the vibrational t-vectors, the $\textbf{t}_{ik}$ vectors in Eq.~(\ref{eq:tvib}) that are the derivatives of the body-fixed Cartesian coordinates with respect to the internal coordinates and are calculated in GENIUSH using finite differences. We have studied the effect of the $\zeta$ step size in the two-sided difference formula (that could have been meaningfully used also near the boundaries for the grid types listed in Table~\ref{tab:gridsize}). The default value for the step size is $\zeta=10^{-5}$~a.u. \cite{MaCzCs09}, which we use in our original study, but we have performed computations also with $\zeta=10^{-6}$ a.u. The `NumStep' column of Table~\ref{tab:tests} shows that the effect of this change for the lowest twenty vibrational states is less than $10^{-5}$~\cm.

\paragraph{PES symmetrization}
Wang and Carrington \cite{WaCa21} pointed out that there is a small imperfection in the AOSz05 PES \cite{AOSz05} with respect to the (123) and (132) permutations of the methane protons, and for this reason they used a symmetrized version of the PES by averaging over the rotated geometries (perfect numerical symmetry was critical for them for using the symmetry-adapted Lanczos eigensolver). The `PESsym' column of Table~\ref{tab:tests} shows that the effect of this operation is less than $3\times 10^{-7}$ \cm.

\paragraph{Generation of the $\bos{G}$ matrix over the grid points by increased precision computer algebra}
To check the numerical KEO procedure for this example, we generated the $\bos{G}$ matrix values over the entire grid using the Wolfram Mathematica symbolic algebra program \cite{WolframMath} with 20 digits precision. Note that we use quadruple precision in the Fortran implementation of GENIUSH for the finite difference calculation of the vibrational t-vectors, but only double precision for the inversion of the $\bos{g}$ matrix. Since near the singularities, we have to deal with small (large) values, and for this reason, we have decided to check the calculation procedure.
The `CompAlg' column of Table~\ref{tab:tests} shows that using an increased precision Mathematica calculation to generate all KEO coefficients has an effect smaller than $10^{-5}$~\cm\ on the first twenty vibrational states of methane-water.

\begin{table}%[htbp]
\caption{%
   Testing the computational setup in GENIUSH for the first twenty vibrational ($J=0$) energy levels, in~\cm, of MW and using the AOSz05 PES. The first energy value corresponds to the zero-point vibrational energy (ZPVE) and the other values are relative to the ZPVE. 
  \label{tab:tests}  
}
\begin{tabular}{@{}r@{\ \ } r@{\ \ }r@{\ \ } c@{\ \ }  c r@{\ \ } c@{\ \ }  c r@{\ \ }c@{\ \ }}
\hline\\[-0.35cm]
\hline\\[-0.35cm]
J0.$n$ & \multicolumn{1}{c}{$\tilde{\nu}_0$}	&
\multicolumn{1}{c}{NumStep$^\text{a}$} &
\multicolumn{1}{c}{PESsym$^\text{b}$} &
\multicolumn{1}{c}{CompAlg$^\text{c}$}   \\
%
%\cline{3-4} \cline{6-7} \cline{9-10} \\[-0.35cm]
%
\hline\\[-0.35cm]
1	 &	206.81021	     &	     $4\times10^{-6}$	    &	   $1\times10^{-8}$	    &	     $1\times10^{-5}$	     \\
2	 &	4.76383	     &	     $4\times10^{-6}$	    &	   $3\times10^{-7}$	    &	     $7\times10^{-6}$	     \\
3	 &	4.76400	     &	     $4\times10^{-6}$	    &	   $3\times10^{-7}$	    &	     $8\times10^{-6}$	     \\
4	 &	4.76464	     &	     $1\times10^{-5}$	    &	   $3\times10^{-8}$	    &	     $9\times10^{-6}$	     \\
5	 &	6.99193	     &	     $3\times10^{-7}$	    &	   $2\times10^{-8}$	    &	     $5\times10^{-6}$	     \\
6	 &	11.25042	     &	     $4\times10^{-6}$	    &	   $3\times10^{-7}$	    &	     $1\times10^{-5}$	     \\
7	 &	11.25051	     &	     $4\times10^{-6}$	    &	   $4\times10^{-7}$	    &	     $1\times10^{-5}$	     \\
8	 &	11.25122	     &	     $1\times10^{-5}$	    &	   $1\times10^{-8}$	    &	     $4\times10^{-6}$	     \\
9	 &	29.03343	     &	     $2\times10^{-6}$	    &	   $6\times10^{-8}$	    &	     $2\times10^{-6}$	     \\
10	 &	29.03361	     &	     $2\times10^{-6}$	    &	   $5\times10^{-8}$	    &	     $6\times10^{-6}$	     \\
11	 &	29.03361	     &	     $2\times10^{-6}$	    &	   $7\times10^{-8}$	    &	     $6\times10^{-6}$	     \\
12	 &	32.63602	     &	     $4\times10^{-6}$	    &	   $3\times10^{-8}$	    &	     $2\times10^{-5}$	     \\
13	 &	32.63712	     &	     $3\times10^{-6}$	    &	   $2\times10^{-7}$	    &	     $1\times10^{-5}$	     \\
14	 &	32.63714	     &	     $3\times10^{-6}$	    &	   $1\times10^{-7}$	    &	     $9\times10^{-6}$	     \\
15	 &	32.71092	     &	     $4\times10^{-6}$	    &	   $2\times10^{-8}$	    &	     $2\times10^{-5}$	     \\
16	 &	32.7121	     &	     $4\times10^{-6}$	    &	   $2\times10^{-7}$	    &	     $8\times10^{-6}$	     \\
17	 &	32.7123	     &	     $4\times10^{-6}$	    &	   $2\times10^{-7}$	    &	     $8\times10^{-6}$	     \\
18	 &	34.41043	     &	     $4\times10^{-6}$	    &	   $6\times10^{-8}$	    &	     $1\times10^{-6}$	     \\
19	 &	35.91968	     &	     $2\times10^{-6}$	    &	   $2\times10^{-8}$	    &	     $1\times10^{-5}$	     \\
20	 &	35.91972	     &	     $2\times10^{-6}$	    &	   $2\times10^{-8}$	    &	     $1\times10^{-5}$	     \\
\hline\\[-0.35cm]
\hline\\[-0.35cm]
\end{tabular}
\begin{flushleft}
$^\text{a}$~$\tilde\nu-\tilde\nu_\text{NumStep}$, where $\tilde\nu_\text{NumStep}$ was obtained by changing the step size of the numerical finite difference calculation of the vibrational t-vectors in GENIUSH from $\zeta = 10^{-5}$~a.u. ($\tilde\nu_0)$ to $\zeta = 10^{-6}$~a.u. (using quadruple precision in Fortran).\\
$^\text{b}$~$\tilde\nu-\tilde\nu_\text{PESsym}$, where $\tilde\nu_\text{PESsym}$ was obtained by  by averaging for every PES points the effect of the identity and the (123) and (132) methane permutation.  \\
$^\text{c}$~$\tilde\nu-\tilde\nu_\text{CompAlg}$, where $\tilde\nu_\text{CompAlg}$ was obtained
by generating the $\bos{G}$ matrix elements using a Wolfram Mathematica implementation with 20 digits precision over the entire grid and performing the vibrational calculations with GENIUSH using these KEO coefficients. \\
\end{flushleft}
\end{table}

%%%%%%%%%%%%%%%%%%%%%%%%%%%%%%%%%%%%%%%%%%%%%%%%%%%%%%%%%%%%%%%%%%%%%%%%%%%%%%%%%%%%%%%%%%%%%
%
% Comparison with Tucker
%
%%%%%%%%%%%%%%%%%%%%%%%%%%%%%%%%%%%%%%%%%%%%%%%%%%%%%%%%%%%%%%%%%%%%%%%%%%%%%%%%%%%%%%%%%%%%%
\clearpage
\section{Comparison with Wigner $D$ basis function computations \label{ch:comparison}}
\noindent 
The dimer Hamiltonian \cite{BrAvSuTe83} used by Wang and Carrington in Ref.~\cite{WaCa21} corresponds to a different coordinate choice from ours. It relies on using 
two full sets of Euler angles (6 angles) to describe the monomers' rotation with respect to the dimer fixed frame (DF), and two additional angles are used to describe the rotational motion, while the separation of the centers of mass of the monomers is described with the $R$ distance, similarly to our work. 
For this angular representation, the KEO can be written in terms of angular momentum operators
and the kinetic energy matrix elements can be calculated analytically using Wigner's $D$ functions \cite{BrAvSuTe83}. 
The Wigner $D$ functions are non-direct product functions, and they efficiently account for the singularities in the KEO. Wang and Carrington \cite{WaCa21} combined this method with the Symmetry Adapted Lanczos (SAL) algorithm to obtain symmetry labels and to make their computations even more efficient.

Table \ref{tab:comparison} presents the comparison of three approaches: 
a) the analytic dimer Hamiltonian and non-direct product Wigner $D$ basis functions of Wang and Carrington \cite{WaCa21};
b) GENIUSH with a numerical KEO and the `smaller' direct-product grid using Legendre polynomials for the singular $\cos\theta$ and $\cos\beta$ coordinates of Ref.~\cite{SaCsAlWaMa16};  
and 
c) the present GENIUSH computations with a numerical KEO and a direct-product grid in which the Legendre DVR is replaced with sine-cot-DVR for the singular coordinates. 

By replacing the Legendre DVRs with sine-cot-DVRs, we can reduce the number of basis functions from 60 million (of the `small' basis \& grid \cite{SaCsAlWaMa16} ) to 13.4 million while the 
the convergence from `error' is reduced from $\sim 1$~\cm\ to $10^{-3}$~\cm. 
Our `optimal' basis with 13.4 million functions are four-five times larger than the Wigner $D$ basis including 2.97 million functions of Ref.~\cite{WaCa21}, but we note that our smaller basis set with 6 million functions in Table~\ref{tab:gridsize} has an convergence error of only on the order of $\pm 3\cdot 10^{-3}$~\cm. 
We also note that we observe a 0.02~\cm\ deviation between our `optimal' basis results (Table~\ref{tab:comparison}) and results of Wang and Carrington~\cite{WaCa21}
for the 36th vibrational state that can be assigned as the stretching fundamental. 
This convergence eror reappears also for $J=1$ (J1.86 in Table~\ref{tab:comparison}) that appears to be the rotational excitation of the stretching fundamental vibration.
Since we have carefully checked the convergence (and other computational parameters) in the present work, we think that this small difference in the energy list would probably disappear if Wang and Carrington \cite{WaCa21} used a slightly larger basis set and grid along the $R$ intermolecular stretching coordinate.

In our computations (DVR), the number of grid points equals the number of basis functions. Wang and Carrington reported only their basis set size, but we have not found any mention of the size of their grid (probably larger than the number of basis functions), while the grid size determines the computational cost (calculation of the matrix-vector product with the potential energy).

All in all, the agreement is remarkable, given the different coordinate representations, the entirely different KEOs and basis sets used to build the Hamiltonian matrix. 
We have not accounted for the symmetry, but the symmetry-adapted Lanczos algorithm can be used also in a GENIUSH computation \cite{FaQuCs17}.
The tailor-made, non-product Wigner $D$ basis function computation remains to be the most efficient apprach, but the numerical performance of the GENIUSH program (using the sine-cot-DVR of Schiffel and Manthe for the singular bending coordinates) becomes comparable to the specialized approach in terms of the basis set size and also in terms of the convergence of the energies.

\begin{table}[htbp]
\caption{%
  Comparison of a representative set of rovibrational energies referenced to the zero-point energy, in \cm, obtained with non-product (NP) and direct-product (DP) basis computations for the methane-water dimer. (The full list is provided in the \som.)
  \label{tab:comparison}  
}
\begin{tabular}{@{}l@{\ \ \ \ } c@{\ \ } c@{\ \ } c@{\ \ } c@{\ \ \ } c@{\ } c@{\ \ } }
\hline\\[-0.35cm]
\hline\\[-0.35cm]
Label & 
Sym.$^\text{a}$ &
\multicolumn{1}{c}{$\tilde\nu_\text{NP}$ \cite{WaCa21}}	&&
\multicolumn{1}{c}{$\tilde\nu'_\text{DP}$ \cite{SaCsAlWaMa16}}&
\multicolumn{1}{c}{$\tilde\nu_\text{DP}$ [This work]}	&
\multicolumn{1}{c}{$\tilde\nu_\text{NP}-\tilde\nu_\text{DP}$} \\
$N_\text{b}^\text{b}$: & &
$3.0\cdot 10^6$	&&
$6.0\cdot 10^7$ &
$1.4\cdot 10^7$	&
 \\

\hline\\[-0.35cm]
J0.1    & A$_1^+ $  &  206.810 &&      206.801 &      206.810 &    0.000   \\[0.1cm]
J0.2    &           &  4.765   &&      4.763   &      4.764   &    0.001   \\
J0.3    & F$_2^+ $  &  4.765   &&      4.764   &      4.764   &    0.001   \\
J0.4    &           &  4.765   &&      4.764   &      4.765   &    0.000   \\[0.1cm]
J0.5    & A$_2^- $  &  6.993   &&      6.934   &      6.992   &    0.001   \\[0.1cm]
J0.18   & A$_1^+ $  &  34.413  &&      34.405  &      34.410  &    0.003   \\[0.1cm]
J0.19   & \multirow{2}{0.4cm}{E$^+$}  &   35.920  &&      35.880  &      35.920  & 0.000 \\
J0.20   &                             &   35.920  &&      35.880  &      35.920  & 0.000 \\[0.1cm]
J0.21   & \multirow{2}{0.4cm}{E$^-$}  &   36.404  &&      36.397  &      36.404  & 0.000 \\
J0.22   &           &  36.404  &&      36.397  &      36.405  & 0.000   \\[0.1cm]
J0.23   &           &  36.414  &&      36.317  &      36.412  & 0.002   \\
J0.24   & F$_1^+ $  &  36.414  &&      36.317  &      36.413  & 0.001   \\
J0.25   &           &  36.414  &&      36.322  &      36.414  & 0.000   \\[0.1cm]
J0.36   & A$_1^+ $  &  48.706  &&      48.685  &      48.686  & {0.020}   \\[0.1cm]
J0.66   & A$_2^- $  &  66.597  &&      66.156  &      66.598  & 0.001   \\[0.1cm]
J1.1    & A$_2^+ $  &  0.289   &&      0.289   &      0.289   & 0.000   \\[0.1cm]
J1.2    &           &  5.047   &&      5.044   &      5.045   & 0.001   \\
J1.3    & F$_1^+ $  &  5.047   &&      5.044   &      5.046   & 0.001   \\
J1.4    &           &  5.047   &&      5.045   &      5.046   & 0.001   \\[0.1cm]
J1.5    & A$_1^- $  &  7.282   &&      7.219   &      7.281   & 0.001   \\[0.1cm]
J1.38   &           & 30.687   &&      30.474  &      30.687  & 0.001   \\
J1.39   & F$_1^+ $  & 30.687   &&      30.474  &      30.687  & 0.001   \\
J1.40   &           & 30.687   &&      30.588  &      30.687  & 0.001   \\[0.1cm]
J1.41   &           & 30.688   &&      30.588  &      30.687  & 0.001   \\
J1.42   & F$_2^+ $  & 30.688   &&      30.633  &      30.687  & 0.001   \\
J1.43   &           & 30.688   &&      30.633  &      30.687  & 0.001   \\[0.1cm]
J1.86   & A$_2^+ $  & 48.981   &&      48.946  &      48.961  & {0.021} \\
\hline\\[-0.35cm]
\hline
\end{tabular}
\begin{flushleft}
$^\text{a}$~Symmetry labels corresponding to the character table of Wang and Carrington \cite{WaCa21} instead of the table originally proposed by Dore et al. \cite{DoSa94} used in Refs.~\cite{SaCsAlWaMa16,SaCsMa17}. \\
$^\text{b}$~Number of basis functions that equals the number of grid points for the DP computations. \\
\end{flushleft}
\end{table}

%%%%%%%%%%%%%%%%%%%%%%%%%%%%%%%%%%%%%%%%%%%%%%%%%%%%%%%%%%%%%%%%%%%%%%%%%%%%%%%%%%%%%%%%%%%%%
%
% Line strength
%
%%%%%%%%%%%%%%%%%%%%%%%%%%%%%%%%%%%%%%%%%%%%%%%%%%%%%%%%%%%%%%%%%%%%%%%%%%%%%%%%%%%%%%%%%%%%%
\clearpage

\begin{figure}
    \begin{center}
      \includegraphics[scale=0.6]{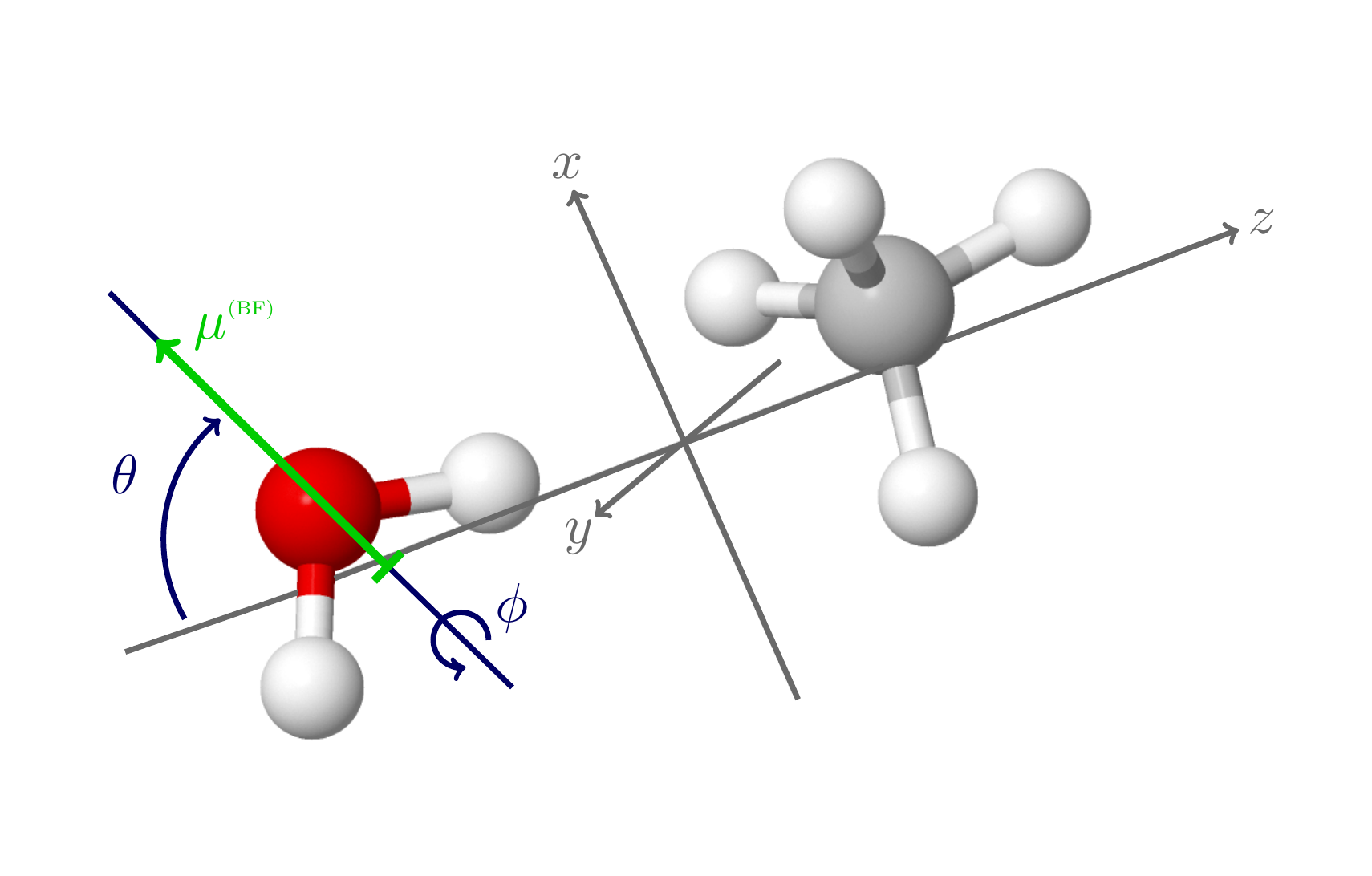}
    \end{center}      
    \caption{
      Schematic representation of the dipole moment in the molecule-fixed frame of the CH$_4$--H$_2$O dimer.
      }
    \label{fig:dipole}
\end{figure}

\section{Evaluation of the line strengths \label{ch:linestr}}
\noindent 
Wang and Carrington \cite{WaCa21} reported also line strength values for the rovibrational transitions using a simple but excellent model for the dimer's dipole moment. They set the dipole moment for the rigid water molecule to 1 (in arbitrary or `normalized, relative' units), since the rigid methane fragment is apolar and does not contribute to this quantity. Using this dipole moment representation, they have computed the line strength for the rovibrational transitions and predicted that the global minimum to the secondary minimum transition should be well visible in the far infrared spectrum.
The transition moment or line strength is defined by \cite{YuBaYa09}
\begin{equation}
\label{eq:linestre}
    S(J'l' \xleftarrow{} Jl) = \text{g}_{\text{ns}} 
    \sum_{m,m'} \sum_{A=X,Y,Z}
    \bigg| \bra{\Psi^{(\text{rv})}_{J'm'l'}} \mu_A 
    \ket{\Psi^{(\text{rv})}_{Jml}} \bigg|^2,
\end{equation}
where $\text{g}_{\text{ns}}$ is the nuclear spin statistical weight factor and $\mu_A$ ($A=X,Y,Z$) are the components of the molecular dipole moment in the laboratory-fixed frame. The $g_\text{ns}$ values have been calculated in Ref.~\cite{SaCsAlWaMa16}. We note that 
Ref.~\cite{WaCa21} reported the `bare' values of the integrals (without the $g_\text{ns}$ factor), that will be labelled as $S_0=S/g_\text{ns}$ in the tables.

The  rovibrational wave functions of the isolated molecule Hamiltonian, Eq.~(\ref{eq:hamil}), are represented in the GENIUSH program \cite{FaMaCs11} as 
\begin{align}
  \ket{\Psi^{(\text{rv})}_{Jml}} 
  = 
  \sum_{v=1}^{N_\text{b}}
  \sum_{k,\tau} 
    c_{v,k,\tau}^{(Jl)} \ket{v} \ket{J,k,m,\tau}
\end{align}
expressed with the $\ket{v}$ vibrational wave function and the Wang-type symmetric top functions $\ket{J,k,m,\tau} = d_{+k}^{(\tau)} \ket{J,k,m} + d_{-k}^{(\tau)} \ket{J,-k,m}$.

\begin{table}%[htbp]
\caption{%
$U^{(\Omega)}_{\omega\sigma,\alpha}$ matrix elements for $\Omega=1$. 
  \label{tab:u}  
}
\begin{tabular}{ c@{\ \ \ } c@{\ \ \ } c@{\ \ \ } c@{\ \ \ } c@{\ \ \ } c@{\ \ \ } c@{\ \ \ }}
\hline\\[-0.35cm]
\hline\\[-0.35cm]
($\omega, \sigma$) &$\alpha$:& $x$ && $y$ && $z$ \\
\hline\\[-0.30cm]
(1,$-$1)  && $\frac{1}{\sqrt{2}}$ && $-\frac{\text{i}}{\sqrt{2}}$ && 0 \\[0.35cm]
(1,0)   && 0 && 0 && 1 \\[0.35cm]
(1,1)   && $-\frac{1}{\sqrt{2}}$ && $-\frac{\text{i}}{\sqrt{2}}$ && 0 \\
\hline\\[-0.35cm]
\hline\\[-0.35cm]
\end{tabular}
\begin{flushleft}
\end{flushleft}
\end{table}
Following Owens and Yachmenev, we evaluate the rovibrational integrals for $\Omega$-order tensorial properties in the laboratory-fixed frame (LF) \cite{OwYa18} as:
 \begin{equation}
 \label{eq:tensmatrix}
    \bra{\Psi^{(\text{rv})}_{J'm'l'}} T^{(\text{LF})}_A 
    \ket{\Psi^{(\text{rv})}_{Jml}} = \sum_{\omega=0}^\Omega
    \mathcal{M}_{A\omega}^{(J'm',Jm)}
    \mathcal{K}_{\omega}^{(J'l',Jl)}
 \end{equation}
with
 \begin{equation}
    \mathcal{M}_{A\omega}^{(J'm',Jm)} = (-1)^{m'}
     \sqrt{(2J'+1)(2J+1)} \times
     \sum_{\sigma=-\omega}^\omega 
     [U^{(\Omega)}]^{-1}_{A,\omega\sigma}
      \begin{pmatrix}
          J & \omega & J'\\
          m & \sigma & -m'
      \end{pmatrix}
 \label{eq:tensmatrix1}
 \end{equation}
and
 \begin{equation}
 \begin{split}
    & \mathcal{K}_{\omega}^{(J'l',Jl)} = 
    \sum_{\substack{v,k,\tau\\v',k',\tau'}}
    [c_{v'k'\tau'}^{(J'l')}]^* c_{vk\tau}^{(Jl)}
     \sum_{\pm k', \pm k}[d_{k'}^{(\tau')}]^*
     d_k^{(\tau)}\\
     & \times (-1)^{k'}
     \sum_{\sigma=-\omega}^\omega \sum_{\alpha} 
      \begin{pmatrix}
          J & \omega & J'\\
          k & \sigma & -k'
      \end{pmatrix}
     \times U^{(\Omega)}_{\omega\sigma, \alpha}
     \bra{v'}T_\alpha^{(\text{BF})}\ket{v}.
 \end{split}
  \label{eq:tensmatrix2}
 \end{equation}
The electric dipole moment is a rank-1 tensor (a vector) with $\Omega=1$,
and the corresponding $U$ matrix values are summarized in Table~\ref{tab:u}.
For higher-order tensor operators, this matrix can be calculated from lower rank tensors using the following expression
\begin{equation}
\label{eq:u}
    U^{(\Omega)}_{\omega\sigma, A} = 
    \sum_{\sigma_1=-\omega_1}^{\omega_1}
    \sum_{\sigma_2=-\omega_2}^{\omega_2}
    \bra{\omega_1\sigma_1\omega_2\sigma_2} \ket{\omega\sigma}
    U^{(\Omega_1)}_{\omega_1\sigma_1, B}  
    U^{(\Omega_2)}_{\omega_2\sigma_2, C},
\end{equation}
where $\Omega=\Omega_1+\Omega_2$, $A=B\otimes C$ and $\bra{\omega_1\sigma_1\omega_2\sigma_2} \ket{\omega\sigma}$ is the Clebsch--Gordan coefficient.

Regarding the transition dipoles of methane-water,
we have first computed the vibrational matrix elements $\bra{v'}\mu_\alpha^{(\text{BF})}\ket{v}$, using the H$_2$O dipole approximation
of Ref.~\cite{WaCa21}, \emph{i.e.,} that only the water monomer contributes to the dipole moment of the dimer (Fig.~\ref{fig:dipole}). 
It is shown in the Figure that in our representation this dipole vector points along the $Z$ axis of the H$_2$O monomer frame, and in the dimer's frame its rotation is described with the $\theta$ and $\phi$ angles (although the value of $\phi$ is irrelevant in this special case).  
Hence, it was straightforward to implement this (presumably excellent) approximation for the body-(dimer-)fixed dipole moment $\mu_\alpha^{(\text{BF})}$. and calculate its value at a grid point in the body-fixed (dimer's) frame, and then integrate it for pairs of vibrational eigenvectors.

%%%%%%%%%%%%%%%%%%%%%%%%%%%%%%%%%%%%%%%%%%%%%%%%%%%%%%%%%%%%%%%%%%%%%%%%%%%%%%%%%%%%%%%%%%%%%
%
% FIR
%
%%%%%%%%%%%%%%%%%%%%%%%%%%%%%%%%%%%%%%%%%%%%%%%%%%%%%%%%%%%%%%%%%%%%%%%%%%%%%%%%%%%%%%%%%%%%%
\clearpage
\section{Far infrared spectrum \label{ch:firspectrum}}
\noindent
The rovibrational transitions computed in this work are compared in Tables~\ref{tab:E1E2}--\ref{tab:AF8} with the  transitions computed by Wang and Carrington \cite{WaCa21} and with the transitions observed experimentally  in the far-infrared range by Dore $et\ al.$ \cite{DoSa94}. 
In addition to the 13 experimentally observed vibration-rotation-tunnelling bands reported by Dore $et\ al.$: $(E)_1$, $(E)_2$, $(A/F)_1$, $(A/F)_2$, $(A/F)_3$ and $(A/F)_4$ for $ortho$-H$_2$O, $(E)_3$, $(E)_4$, $(E)_5$, $(A/F)_5$, $(A/F)_6$, $(A/F)_7$ and $(A/F)_8$ for $para$-H$_2$O; 
Wang and Carrington predicted 8 additional bands \cite{WaCa21} (and of course computations can predict more): $(A/F)_{1b}$, $(A/F)_{3X}$, $(A/F)_{4X}$, $(A/F)_{5X}$, $(A/F)_{6X}$, $(A/F)_{6Y}$, $(A/F)_{8X}$ and $(A/F)_{8Y}$. 
The tables show that the two computations are in an excellent agreement. regarding the theory-experiment comparison, we reiterate the early observation \cite{SaCsAlWaMa16}, according to which the discrepancies are larger for the transitions including the $para$-H$_2$O than the $ortho$-H$_2$O species. 

\begin{table}[htbp]
\caption{%
$(E)_1$ and $(E)_2$ bands of the CH$_4$--H$_2$O. 
All transitions are reported in \cm\ except where it is otherwise stated. 
$(J,n)'$ and $(J,n)$ label the rovibrational state in our energy lists provided in the \som.
[this w]: [this work]. 
$\Delta\nu=E'-E$. 
$S_0$ is the line strength, Eq.~(\ref{eq:linestre}), without the spin statistical weight factor \cite{SaCsAlWaMa16}.
\label{tab:E1E2}  
}
\small{
\begin{tabular}{@{}l@{\ \ } c@{\ \ }c@{\ \ \ }  c@{\ \ } c@{\ \ \ } c@{\ \ } c@{\ \ }c@{\ \ } c@{}}
\hline\\[-0.35cm]
\hline\\[-0.35cm]
$(J,n)'$ $\xleftarrow{}$  $(J,n)$ &
$\Delta\tilde\nu_\text{obs}$~[MHz] &  $\Delta\tilde\nu_\text{obs}$  &
$E'$ & $E$ 
& $\Delta\tilde\nu$ & $\Delta\tilde\nu$ &
$S_0$ & 
$S_0$ \\
\\[-0.35cm]
&
\cite{DoSa94} & &
[this w]& [this w]& [this w] & \cite{WaCa21} &
[this w] & 
\cite{WaCa21} \\
\\[-0.35cm]
\hline\\[-0.35cm]
& \multicolumn{8}{l}{Expt.~\cite{DoSa94}: Table III, $(E)_1$, $\Sigma \xleftarrow{} \Pi $ band} \\
0, 21--22        $\xleftarrow{}$ 1, 23--24        &       532812.0        &       17.7727 &       36.4045 &       19.3909 &       17.0136 &       17.0099 &       0.155   &       0.154   \\
1, 53--54        $\xleftarrow{}$ 1, 23--24        &       541359.5        &       18.0578 &       36.6889 &       19.3909 &       17.2980 &       17.2946 &       0.232   &       0.230   \\
2, 75--76        $\xleftarrow{}$ 1, 23--24        &       558451.2        &       18.6279 &       37.2577 &       19.3909 &       17.8668        &       17.8638 &      0.074         &       0.073   \\
1, 53--54        $\xleftarrow{}$ 2, 23--24        &       524254.1        &       17.4872 &       36.6889 &       19.9581 &       16.7309 &       16.7270 &       0.232   &       0.230   \\
2, 75--76        $\xleftarrow{}$ 2, 23--24        &       541344.3        &       18.0573 &       37.2577 &       19.9581 &       17.2996 &       17.2962 &           0.393    &       0.391   \\
\\
& \multicolumn{8}{l}{Expt.~\cite{DoSa94}: Table III, $(E)_2$, $\Delta \xleftarrow{} \Pi $ band} \\
2,	80--81 $\xleftarrow{}$	2,	23--24	&	542744.8	&	18.1040	&	37.5721	&	19.9581	&	17.6140	&	17.6100	&	0.263	&	0.262	\\
2,	80--81 $\xleftarrow{}$	1,	23--24	&	559850.8	&	18.6746	&	37.5721	&	19.3909	&	18.1812	&	18.1800	&	0.484	&	0.482	\\

\hline\\[-0.35cm]
\hline\\[-0.35cm]
\end{tabular}

\begin{flushleft}
%
%$^\text{a}$~
%
\end{flushleft}

}
\end{table}

\begin{table}[htbp]
\caption{%
$(A/F)_1$ and $(A/F)_3$ bands of the CH$_4$--H$_2$O. 
See also caption to Table~\ref{tab:E1E2}.
  \label{tab:AF1AF3}  
}
\small{
\begin{tabular}{@{}l@{\ \ } c@{\ \ }c@{\ \ \ }  c@{\ \ } c@{\ \ \ } c@{\ \ } c@{\ \ }c@{\ \ } c@{}}
\hline\\[-0.35cm]
\hline\\[-0.35cm]
$(J,n)'$ $\xleftarrow{}$  $(J,n)$ &
$\Delta\tilde\nu_\text{obs}$~[MHz] &  $\Delta\tilde\nu_\text{obs}$  &
$E'$ & $E$ 
& $\Delta\tilde\nu$ & $\Delta\tilde\nu$ &
$S_0$ & 
$S_0$ \\
\\[-0.35cm]
&
\cite{DoSa94} & &
[this w]& [this w]& [this w] & \cite{WaCa21} &
[this w] & 
\cite{WaCa21} \\
\\[-0.35cm]
\hline\\[-0.35cm]
& \multicolumn{8}{l}{Expt. (Ref. \cite{DoSa94}): Table IV, $(A/F)_1$, $\Sigma \xleftarrow{} \Pi $ band} \\
0, 12--14        $\xleftarrow{}$ 1, 20--22        &       538189.8        &       17.9521 &       32.6368 &       14.6152 &       18.0215 &       18.0194 &       0.167   &       0.166   \\
1, 47--49        $\xleftarrow{}$ 1, 20--22        &       546831.5        &       18.2403 &       33.0018 &       14.6152 &       18.3866 &       18.3839 &       0.231   &       0.230   \\
2, 60--62        $\xleftarrow{}$ 1, 20--22        &       564291.7        &       18.8227 &       33.5026 &       14.6152 &       18.8874 &       18.8858 &       0.080        &       0.080   \\
1, 47--49        $\xleftarrow{}$ 2, 17--19        &       529495.8        &       17.6621 &       33.0018 &       15.1903 &       17.8115 &       17.8083 &       0.234   &       0.232   \\
2, 60--62        $\xleftarrow{}$ 2, 17--19        &       546715.2        &       18.2365 &       33.5026 &       15.1903 &       18.3123 &       18.3102 &       0.420        &       0.417   \\
\\
& \multicolumn{8}{l}{Predicted in Ref.~\cite{WaCa21}: $(A/F)_{1b}$, $\Delta \xleftarrow{} \Pi $ band} \\
0, 15--17        $\xleftarrow{}$ 1, 17--19        &       --       &       --       &       32.7118 &       14.6124 &       18.0994 &       18.0960 &       0.155   &       0.154   \\
1, 44--46        $\xleftarrow{}$ 1, 17--19        &       --       &       --       &       32.9254 &       14.6124 &       18.3130 &       18.3106 &       0.249   &       0.248   \\
2, 63--65        $\xleftarrow{}$ 1, 17--19        &       --       &       --       &       33.5818 &       14.6124 &       18.9693 &       18.9667 &       0.076        &       0.076   \\
1, 44--46        $\xleftarrow{}$ 2, 20--22        &       --       &       --       &       32.9254 &       15.1975 &       17.7279 &       17.7254 &       0.251   &       0.250   \\
2, 63--65        $\xleftarrow{}$ 2, 20--22        &       --       &       --       &       33.5818 &       15.1975 &       18.3843 &       18.3815 &       0.387        &       0.384   \\
\\
& \multicolumn{8}{l}{Expt. (Ref. \cite{DoSa94}): Table IV, $(A/F)_{3a}$, $\Delta \xleftarrow{} \Pi $ band} \\
2, {66--71}$^\text{a}$        $\xleftarrow{}$ 2, 17--19        &       564637.3        &       18.8343 &          34.1923     &       15.1903 &   19.0020    &       18.9991 &        0.217       &       0.215   \\
2, {66--71}$^\text{a}$         $\xleftarrow{}$ 1, 20--22        &       581971.4        &       19.4125 &         34.1923      &       14.6152 &  19.5770     &       19.5747 &       0.392        &       0.389   \\
\\
& \multicolumn{8}{l}{Expt. (Ref. \cite{DoSa94}): Table IV, $(A/F)_{3b}$, $\Delta \xleftarrow{} \Pi $ band} \\
2, {66--71}$^\text{a}$        $\xleftarrow{}$ 2, 20--22        &       564437.7        &       18.8276 &      34.1925   &       15.1975 &     18.9950  &       18.9919 &      0.218         &       0.217   \\
2, {66--71}$^\text{a}$        $\xleftarrow{}$ 1, 17--19        &       582013.5        &       19.4139 &      34.1925   &       14.6124 &     19.5801 &       19.5771 &       0.391        &       0.388   \\
\\
& \multicolumn{8}{l}{Predicted in Ref.~\cite{WaCa21}: $(A/F)_{3Xa}$, $\Delta \xleftarrow{} \Pi $ band} \\
2, {54--59}$^\text{a}$        $\xleftarrow{}$ 2, 17--19        &       --       &       --       &       32.8319 &       15.1903 &      17.6416  &       17.6405 &      0.040         &       0.040   \\
2, {54--59}$^\text{a}$        $\xleftarrow{}$ 1, 20--22        &       --       &       --       &       32.8319 &       14.6152 &        18.2167  &       18.2161 &        0.075       &       0.075   \\
\\
& \multicolumn{8}{l}{Predicted in Ref.~\cite{WaCa21}: $(A/F)_{3Xb}$, $\Delta \xleftarrow{} \Pi$ band} \\
2, {54--59}$^\text{a}$        $\xleftarrow{}$ 2, 20--22        &       --       &       --       &      32.8331  &       15.1975 &        17.6357  &       17.6333 &        0.042       &       0.042$^\text{b}$   \\
2, {54--59}$^\text{a}$        $\xleftarrow{}$ 1, 17--19        &       --       &       --       &      32.8331  &       14.6124 &        18.2207  &       18.2185 &        0.074       &       0.074$^\text{b}$   \\
\hline\\[-0.35cm]
\hline
\end{tabular}
\begin{flushleft}
$^\text{a}$~Only three of the six upper states, which are very close in energy (and for this reason, listed together in our tables), give contribution to the $S_0$ line strength  \\
$^\text{b}$: These two values are interchanged in Ref.~\cite{WaCa21}. \\
\end{flushleft}

}
\end{table}

\begin{table}[htbp]
\caption{%
$(A/F)_2$ band of the CH$_4$-H$_2$O. 
See also caption to Table~\ref{tab:E1E2}.
  \label{tab:AF2}  
}
\small{
\begin{tabular}{@{}l@{\ \ } c@{\ \ }c@{\ \ \ }  c@{\ \ } c@{\ \ \ } c@{\ \ } c@{\ \ }c@{\ \ } c@{}}
\hline\\[-0.35cm]
\hline\\[-0.35cm]
$(J,n)'$ $\xleftarrow{}$  $(J,n)$ &
$\Delta\tilde\nu_\text{obs}$~[MHz] &  $\Delta\tilde\nu_\text{obs}$  &
$E'$ & $E$ 
& $\Delta\tilde\nu$ & $\Delta\tilde\nu$ &
$S_0$ & 
$S_0$ \\
\\[-0.35cm]
&
\cite{DoSa94} & &
[this w]& [this w]& [this w] & \cite{WaCa21} &
[this w] & 
\cite{WaCa21} \\
\\[-0.35cm]
\hline\\[-0.35cm]
& \multicolumn{8}{l}{Expt. (Ref. \cite{DoSa94}): Table V, $(A/F)_2$, $\Pi \xleftarrow{} \Sigma $ band} \\
1, 35--37        $\xleftarrow{}$ 0, 6--8  &       562445.5        &       18.7612 &       29.7641 &       11.2507 &       18.5134 &       18.5106 &       0.319   &       0.317   \\
1, 32--34        $\xleftarrow{}$ 1, 12--14        &       553888.4        &       18.4757 &       29.7619 &       11.5324 &       18.2294 &       18.2265 &       0.480   &       0.477   \\
1, 35--37        $\xleftarrow{}$ 2, 12--14        &       536931.0        &       17.9101 &       29.7641 &       12.0959 &       17.6682 &       17.6648 &       0.161   &       0.160   \\
2, 45--47        $\xleftarrow{}$ 1, 12--14        &       571055.5        &       19.0484 &       30.3357 &       11.5324 &   18.8033     &       18.8009 &       0.477        &       0.475   \\
2, 42--44        $\xleftarrow{}$ 2, 12--14        &       553883.6        &       18.4756 &       30.3291 &       12.0959 &   18.2332    &       18.2303 &        0.799       &       0.794   \\

\hline\\[-0.35cm]
\hline\\[-0.35cm]
\end{tabular}

\begin{flushleft}
%
%$^\text{a}$~
%
\end{flushleft}
}
\end{table}

\begin{table}[htbp]
\caption{%
$(A/F)_4$ band of the CH$_4$-H$_2$O. 
See also caption to Table~\ref{tab:E1E2}.
  \label{tab:AF4}  
}
\small{
\begin{tabular}{@{}l@{\ \ } c@{\ \ }c@{\ \ \ }  c@{\ \ } c@{\ \ \ } c@{\ \ } c@{\ \ }c@{\ \ } c@{}}
\hline\\[-0.35cm]
\hline\\[-0.35cm]
$(J,n)'$ $\xleftarrow{}$  $(J,n)$ &
$\Delta\tilde\nu_\text{obs}$~[MHz] &  $\Delta\tilde\nu_\text{obs}$  &
$E'$ & $E$ 
& $\Delta\tilde\nu$ & $\Delta\tilde\nu$ &
$S_0$ & 
$S_0$ \\
\\[-0.35cm]
&
\cite{DoSa94} & &
[this w]& [this w]& [this w] & \cite{WaCa21} &
[this w] & 
\cite{WaCa21} \\
\\[-0.35cm]
\hline\\[-0.35cm]
& \multicolumn{8}{l}{Expt. (Ref. \cite{DoSa94}): Table VI, $(A/F)_4$, $\Pi \xleftarrow{} \Sigma $ band} \\
1, 26   $\xleftarrow{}$ 0, 5    &       574574.9        &       19.1658 &       26.3646 &       6.9919  &       19.3727 &       19.3705 &       0.313   &       0.311   \\
1, 25   $\xleftarrow{}$ 1, 5    &       565794.7        &       18.8729 &       26.3621 &       7.2805  &       19.0816 &       19.0792 &       0.473   &       0.471   \\
1, 26   $\xleftarrow{}$ 2, 5    &       548506.7        &       18.2962 &       26.3646 &       7.8576  &       18.5070 &       18.5043 &       0.160   &       0.158   \\
2, 34   $\xleftarrow{}$ 1, 5    &       583344.0        &       19.4583 &       26.9463 &       7.2805  &       19.6657 &       19.6638 &       0.467        &       0.464   \\
2, 33   $\xleftarrow{}$ 2, 5    &       565694.1        &       18.8695 &       26.9386 &       7.8576  &       19.0810 &       19.0787 &       0.787   &       0.783   \\
\\
& \multicolumn{8}{l}{Predicted in Ref.~\cite{WaCa21}: $(A/F)_{4X}$, $\Pi \xleftarrow{} \Sigma $ band} \\
1, 69   $\xleftarrow{}$ 0, 5    &       --       &       --       &       43.0625 &       6.9919  &       36.0706 &       36.0682 &       0.046   &       0.046   \\
1, 70   $\xleftarrow{}$ 1, 5    &       --       &       --       &       43.0660 &       7.2805  &       35.7855 &       35.7829 &       0.071   &       0.071   \\
1, 69   $\xleftarrow{}$ 2, 5    &       --       &       --       &       43.0625 &       7.8576  &       35.2049 &       35.2020  &       0.024   &       0.024   \\
2, 101  $\xleftarrow{}$ 1, 5    &       --       &       --       &       43.6042 &       7.2805  &          36.3236 &       36.3213 &       0.068   &       0.068   \\
2, 102  $\xleftarrow{}$ 2, 5    &       --       &       --       &       43.6147 &       7.8576  &       35.7571 &       35.7544 &       0.118   &       0.118   \\
\hline\\[-0.35cm]
\hline\\[-0.35cm]
\end{tabular}

\begin{flushleft}
%
%$^\text{a}$~
%
\end{flushleft}
}
\end{table}

\begin{table}[htbp]
\caption{%
$(E)_3$, $(E)_4$ and $(E)_5$ bands of the CH$_4$--H$_2$O. 
See also caption to Table~\ref{tab:E1E2}.
  \label{tab:E3E4E5}  
}
\small{
\begin{tabular}{@{}l@{\ \ } c@{\ \ }c@{\ \ \ }  c@{\ \ } c@{\ \ \ } c@{\ \ } c@{\ \ }c@{\ \ } c@{}}
\hline\\[-0.35cm]
\hline\\[-0.35cm]
$(J,n)'$ $\xleftarrow{}$  $(J,n)$ &
$\Delta\tilde\nu_\text{obs}$~[MHz] &  $\Delta\tilde\nu_\text{obs}$  &
$E'$ & $E$ 
& $\Delta\tilde\nu$ & $\Delta\tilde\nu$ &
$S_0$ & 
$S_0$ \\
\\[-0.35cm]
&
\cite{DoSa94} & &
[this w]& [this w]& [this w] & \cite{WaCa21} &
[this w] & 
\cite{WaCa21} \\
\\[-0.35cm]
\hline\\[-0.35cm]
& \multicolumn{8}{l}{Expt. (Ref. \cite{DoSa94}): Table VII, $(E)_3$, $\Sigma \xleftarrow{} \Pi $ band} \\
0, 19--20        $\xleftarrow{}$ 1, 15--16        &       732385.1        &       24.4297 &       35.9197 &       13.1087 &       22.8110 &       22.8098 &       0.093   &       0.092   \\
1, 51--52        $\xleftarrow{}$ 1, 15--16        &       740778.1        &       24.7097 &       36.1856 &       13.1087 &       23.0769 &       23.0757 &       0.141   &       0.140   \\
1, 51--52        $\xleftarrow{}$ 2, 15--16        &       723657.8        &       24.1386 &       36.1856 &       13.6782 &       23.0769 &       22.5057 &       0.136   &       0.135   \\
2, 73--74        $\xleftarrow{}$ 1, 15--16        &       757561.5        &       25.2695 &       36.7175 &       13.1087 &   23.6088     &       23.6078 &       0.048        &       0.048   \\
2, 73--74        $\xleftarrow{}$ 2, 15--16        &       740443.2        &       24.6985 &       36.7175 &       13.6782 &   23.0393     &       23.0378 &       0.232        &       0.231   \\
\\
& \multicolumn{8}{l}{Expt. (Ref. \cite{DoSa94}): Table VII, $(E)_4$, $\Sigma \xleftarrow{} \Pi $ band} \\
2, 88--89        $\xleftarrow{}$ 2, 15--16        &       821483.9        &       27.4018 &       38.6481 &       13.6782 &   24.9700     &       24.9680  &      0.212        &       0.212   \\
2, 88--89        $\xleftarrow{}$ 1, 15--16        &       838603.8        &       27.9728 &       38.6481 &       13.1087 &   25.5395     &       25.5380  &      0.396        &       0.394   \\
\\
& \multicolumn{8}{l}{Expt. (Ref. \cite{DoSa94}): Table VII, $(E)_4'$, $\Sigma \xleftarrow{} \Pi $ band} \\
2, 90--91        $\xleftarrow{}$ 2, 15--16        &       821483.9        &       27.4018 &       39.5975  &       13.6782 &  25.9193      &       25.9171 &      0.005         &       0.005   \\
2, 90--91        $\xleftarrow{}$ 1, 15--16        &       838603.8        &       27.9728 &       39.5975  &       13.1087 &  26.4888      &       26.4871 &      0.001       &       0.001   \\
{1}, {58--59}*        $\xleftarrow{}$ 2, 15--16        &       --       &       --             &       39.0170 &     13.6782 &    25.3388    &     25.3362$^\text{a}$   &       0.001      &       0.001   \\
{1}, {58--59}*        $\xleftarrow{}$ 1, 15--16        &       --       &       --       &    39.0170           &       13.1087 &     25.9083 &     25.9062$^\text{a}$  &    0.001         &       0.001   \\
\\
& \multicolumn{8}{l}{Expt. (Ref. \cite{DoSa94}): Table VII, $(E)_5$, $\Sigma \xleftarrow{} \Pi $ band} \\
0, 34--35        $\xleftarrow{}$ 1, 15--16        &       --       &       --       &       48.0961 &       13.1087 &       34.9874 &       34.9863 &       0.088   &       0.087   \\
1, 84--85        $\xleftarrow{}$ 1, 15--16        &       1057943.1       &       35.2892 &       48.3769 &       13.1087 &       35.2682 &       35.2673 &       0.132   &       0.132   \\
2, 116--117      $\xleftarrow{}$ 1, 15--16        &       1074920.4       &       35.8555 &       48.9384 &       13.1087 &       35.8297 &       35.8292 &       0.044        &       0.044   \\
2, 116--117      $\xleftarrow{}$ 2, 15--16        &       1057801.7       &       35.2845 &       48.9384 &       13.6782 &       35.2602 &       35.2592 &       0.221        &       0.219   \\
\hline\\[-0.35cm]
\hline
\end{tabular}

\begin{flushleft}
$^\text{a}$~We believe that there were some typos in Table~XVIII in Ref. \cite{WaCa21}. Corrected values and labels are given here in comparison with the values computed in this work.
\end{flushleft}
}
\end{table}

\begin{table}[htbp]
\caption{%
$(A/F)_5$ band of the CH$_4$--H$_2$O. 
See also caption to Table \ref{tab:E1E2}.
  \label{tab:AF5}  
}
\small{
\begin{tabular}{@{}l@{\ \ } c@{\ \ }c@{\ \ \ }  c@{\ \ } c@{\ \ \ } c@{\ \ } c@{\ \ }c@{\ \ } c@{}}
\hline\\[-0.35cm]
\hline\\[-0.35cm]
$(J,n)'$ $\xleftarrow{}$  $(J,n)$ &
$\Delta\tilde\nu_\text{obs}$~[MHz] &  $\Delta\tilde\nu_\text{obs}$  &
$E'$ & $E$ 
& $\Delta\tilde\nu$ & $\Delta\tilde\nu$ &
$S_0$ & 
$S_0$ \\
\\[-0.35cm]
&
\cite{DoSa94} & &
[this w]& [this w]& [this w] & \cite{WaCa21} &
[this w] & 
\cite{WaCa21} \\
\\[-0.35cm]
\hline\\[-0.35cm]
& \multicolumn{8}{l}{Expt. (Ref. \cite{DoSa94}): Table VIII, $(A/F)_5$, $\Pi \xleftarrow{} \Sigma $ band} \\
1, 38--40        $\xleftarrow{}$ 0, 2--4  &       852462.1        &       28.4351 &       30.6867 &       4.7642  &       25.9225 &       25.9223 &       0.256   &       0.255   \\
1, 41--43        $\xleftarrow{}$ 1, 2--4  &       844268.9        &       28.1618 &       30.6873 &       5.0455  &       25.6417 &       25.6413 &       0.368   &       0.368   \\
1, 38--40        $\xleftarrow{}$ 2, 2--4  &       827031.4        &       27.5868 &       30.6867 &       5.6083  &       25.0784 &       25.0774 &       0.114   &       0.113   \\
2, 48--50        $\xleftarrow{}$ 1, 2--4  &       860602.5        &       28.7066 &       31.2548 &       5.0455  &       26.2093 &       26.2092 &       0.399   &       0.397   \\
2, 51--53        $\xleftarrow{}$ 2, 2--4  &       844500.0        &       28.1695 &       31.2568 &       5.6083  &       25.6485 &       25.6479 &       0.615   &       0.612   \\
\\
& \multicolumn{8}{l}{$(A/F)_{5X}$, $\Pi \xleftarrow{} \Sigma $ band} \\
1, 60--62        $\xleftarrow{}$ 0, 2--4  &       --       &       --       &       40.4210 &       4.7642  &       35.6569 &       35.6572 &       0.105   &       0.105   \\
1, 63--64        $\xleftarrow{}$ 1, 2--4  &       --       &       --       &       40.4225 &       5.0455  &       35.3769 &       35.3770 &       0.164   &       0.164   \\
1, 60--62        $\xleftarrow{}$ 2, 2--4  &       --       &       --       &       40.4210 &       5.6083  &       34.8127 &       34.8123 &       0.060   &       0.060   \\
2, 92--94        $\xleftarrow{}$ 1, 2--4  &       --       &       --       &       40.9716 &       5.0455  &       35.9261 &       35.9267 &       0.150   &       0.150   \\
2, 95--97        $\xleftarrow{}$ 2, 2--4  &       --       &       --       &       40.9761 &       5.6083  &       35.3678 &       35.3678 &       0.273   &       0.271   \\
\hline\\[-0.35cm]
\hline\\[-0.35cm]
\end{tabular}

\begin{flushleft}
%
%$^\text{a}$~
%
\end{flushleft}
}
\end{table}

\begin{table}[htbp]
\caption{%
$(A/F)_6$ and $(A/F)_7$ bands of the CH$_4$--H$_2$O. 
See also caption to Table~\ref{tab:E1E2}.
  \label{tab:AF6AF7}  
}
\small{
\begin{tabular}{@{}l@{\ \ } c@{\ \ }c@{\ \ \ }  c@{\ \ } c@{\ \ \ } c@{\ \ } c@{\ \ }c@{\ \ } c@{}}
\hline\\[-0.35cm]
\hline\\[-0.35cm]
$(J,n)'$ $\xleftarrow{}$  $(J,n)$ &
$\Delta\tilde\nu_\text{obs}$~[MHz] &  $\Delta\tilde\nu_\text{obs}$  &
$E'$ & $E$ 
& $\Delta\tilde\nu$ & $\Delta\tilde\nu$ &
$S_0$ & 
$S_0$ \\
\\[-0.35cm]
&
\cite{DoSa94} & &
[this w]& [this w]& [this w] & \cite{WaCa21} &
[this w] & 
\cite{WaCa21} \\
\\[-0.35cm]
\hline\\[-0.35cm]
& \multicolumn{8}{l}{Expt. (Ref. \cite{DoSa94}): Table IX, $(A/F)_6$, $\Sigma \xleftarrow{} \Pi $ band} \\
0, 23--25        $\xleftarrow{}$ 1, 6--8  &       --       &       --       &       36.4131 &       7.9119  &       28.5012 &       28.5020 &       0.181   &       0.180   \\
1, 55--57        $\xleftarrow{}$ 1, 9--11 &       906722.4        &       30.2450 &       36.6984 &       7.9158  &       28.7826 &       28.7837 &       0.276   &       0.275   \\
1, 55--57        $\xleftarrow{}$ 2, 6--8  &       889393.8        &       29.6670 &       36.6984 &       8.4916  &       28.2069 &       28.2075 &       0.265   &       0.264   \\
2, 77--79        $\xleftarrow{}$ 1, 6--8  &       924003.8        &       30.8214 &       37.2689 &       7.9119  &       29.3570 &       29.3583 &       0.103   &       0.103   \\
2, 77--79        $\xleftarrow{}$ 2, 9--11 &       906302.4        &       30.2310 &       37.2689 &       8.5026  &       28.7663 &       28.7672 &       0.446   &       0.444   \\
\\
& \multicolumn{8}{l}{Predicted in Ref.~\cite{WaCa21}: $(A/F)_{6X}$, $\Sigma \xleftarrow{} \Pi $ band} \\
0, 9--11     ~$\xleftarrow{}$ 1, 9--11 &       -       &       -       &       29.0335 &       7.9158  &       21.1177 &       21.118  &       0.080   &       0.080   \\
1, 27--29        $\xleftarrow{}$ 1, 6--8  &       -       &       -       &       29.3049 &       7.9119  &       21.3930 &       21.3563 &       0.117   &       0.117   \\
1, 27--29        $\xleftarrow{}$ 2, 9--11 &       -       &       -       &       29.3049 &       8.5026  &       20.8023 &       20.8022 &       0.123   &       0.122   \\
2, 37--39        $\xleftarrow{}$ 1, 9--11 &       -       &       -       &       29.8476 &       7.9158  &       21.9318 &       21.9325 &       0.038   &       0.037   \\
2, 37--39        $\xleftarrow{}$ 2, 6--8  &       -       &       -       &       29.8476 &       8.4916  &       21.3560 &       21.3563 &       0.194   &       0.193   \\
\\
& \multicolumn{8}{l}{Predicted in Ref.~\cite{WaCa21}: $(A/F)_{6Y}$, $\Sigma \xleftarrow{} \Pi $ band} \\
0, 26--28        $\xleftarrow{}$ 1, 9--11 &       --       &       --       &       41.1844 &       7.9158  &       33.2686 &       33.2700 &       0.101   &       0.100   \\
1, 66--68        $\xleftarrow{}$ 1, 6--8  &       --       &       --       &       41.4639 &       7.9119  &       33.5520 &       33.5535$^\text{a}$ &       0.157   &       0.156   \\
1, 66--68        $\xleftarrow{}$ 2, 9--11 &       --       &       --       &       41.4639 &       8.5026  &       32.9613 &       32.9624 &       0.147   &       0.147   \\
2, 98--100       $\xleftarrow{}$ 1, 9--11 &       --       &       --       &       42.0229 &       7.9158  &       34.1071 &       34.1092 &       0.055   &       0.055   \\
2, 98--100       $\xleftarrow{}$ 2, 6--8  &       --       &       --       &       42..229 &       8.4916  &       33.5313 &       33.5330  &      0.262   &       0.261   \\
\\
& \multicolumn{8}{l}{Expt. (Ref. \cite{DoSa94}): Table IX, $(A/F)_{7a}$, $\Sigma \xleftarrow{} \Pi $ band} \\
2, {82--87}$^\text{b}$        $\xleftarrow{}$ 2, 6--8  &       912803.3        &       30.4478 &       37.6439      &       8.4916  &          29.1523     &       29.1532 &       0.299        &       0.297   \\
2, {82--87}$^\text{b}$        $\xleftarrow{}$ 1, 9--11 &       930130.4        &       31.0258 &       37.6439      &       7.9158  &          29.7281     &       29.7294 &       0.536        &       0.533   \\
\\
& \multicolumn{8}{l}{Expt. (Ref. \cite{DoSa94}): Table IX, $(A/F)_{7b}$, $\Sigma \xleftarrow{} \Pi $ band} \\
2, {82--87}$^\text{b}$        $\xleftarrow{}$ 2, 9--11 &       912529.1        &       30.4387 &       37.6447       &       8.5026  &         29.1421      &       29.1421 &          0.309     &       0.307   \\
2, {82--87}$^\text{b}$        $\xleftarrow{}$ 1, 6--8  &       930230.6        &       31.0292 &       37.6447       &       7.9119  &         29.7328      &       29.7332 &          0.530     &       0.527   \\
\hline\\[-0.35cm]
\hline\\[-0.35cm]
\end{tabular}

\begin{flushleft}
$^\text{a}$~A typo in Table~XX of Ref.~\cite{WaCa21} i corrected based on the $E'$ and $E''$ level energies given in the paper and checked against results of this work. \\
$^\text{b}$~See footnote $a$ to Table~\ref{tab:AF1AF3}.
\end{flushleft}
}
\end{table}

\begin{table}[htbp]
\caption{%
$(A/F)_8$ band of the CH$_4$--H$_2$O. 
See also caption to Table~\ref{tab:E1E2}.
  \label{tab:AF8}  
}
\small{%
\begin{tabular}{@{}l@{\ \ } c@{\ \ }c@{\ \ \ }  c@{\ \ } c@{\ \ \ } c@{\ \ } c@{\ \ }c@{\ \ } c@{}}
\hline\\[-0.35cm]
\hline\\[-0.35cm]
$(J,n)'$ $\xleftarrow{}$  $(J,n)$ &
$\Delta\tilde\nu_\text{obs}$~[MHz] &  $\Delta\tilde\nu_\text{obs}$  &
$E'$ & $E$ 
& $\Delta\tilde\nu$ & $\Delta\tilde\nu$ &
$S_0$ & 
$S_0$ \\
\\[-0.35cm]
&
\cite{DoSa94} & &
[this w]& [this w]& [this w] & \cite{WaCa21} &
[this w] & 
\cite{WaCa21} \\
\\[-0.35cm]
\hline\\[-0.35cm]
& \multicolumn{8}{l}{Expt. (Ref. \cite{DoSa94}): Table X, $(A/F)_8$, $\Pi \xleftarrow{} \Sigma $ band} \\
1, 30   $\xleftarrow{}$ 0, 1    &       927673.3        &       30.9439 &       29.5588 &       0.0000  &       29.5588 &       29.5596 &       0.374   &       0.372   \\
1, 31   $\xleftarrow{}$ 1, 1    &       919166.0        &       30.6601 &       29.5646 &       0.2891  &       29.2755 &       29.2762 &       0.554   &       0.551   \\
1, 30   $\xleftarrow{}$ 2, 1    &       901595.6        &       30.0740 &       29.5588 &       0.8673  &       28.6915 &       28.6918 &       0.179   &       0.178   \\
2, 40   $\xleftarrow{}$ 1, 1    &       936059.3        &       31.2236 &       30.1269 &       0.2891  &       29.8377 &       29.5596 &       0.569   &       0.566   \\
2, 41   $\xleftarrow{}$ 2, 1    &       919232.4        &       30.6623 &       30.1444 &       0.8673  &       29.2771 &       29.2777 &       0.922   &       0.918   \\
\\
& \multicolumn{8}{l}{Predicted in Ref.~\cite{WaCa21}: $(A/F)_{8X}$, $\Pi \xleftarrow{} \Sigma $ band} \\
0, 18   $\xleftarrow{}$ 1, 1    &       --       &       --       &       34.4104 &       0.2891  &       34.1213 &       34.1234 &       0.048   &       0.048   \\
1, 50   $\xleftarrow{}$ 0, 1    &       --       &       --       &       34.6785 &       0.0000  &       34.6785 &       34.6811 &       0.046   &       0.046   \\
1, 50   $\xleftarrow{}$ 2, 1    &       --       &       --       &       34.6785 &       0.8673  &       34.3894 &       33.8133 &       0.097   &       0.096   \\
2, 72   $\xleftarrow{}$ 1, 1    &       --       &       --       &       35.2146 &       0.2891  &           34.9254 &       34.9283 &       0.091   &       0.090   \\
\\
& \multicolumn{8}{l}{Predicted in Ref.~\cite{WaCa21}: $(A/F)_{8Y}$, $\Pi \xleftarrow{} \Sigma $ band} \\
0, 36   $\xleftarrow{}$ 1, 1    &       --       &       --       &       48.6856 &       0.2891  &       48.3965 &       48.4168 &       0.029   &       0.029   \\
1, 86   $\xleftarrow{}$ 0, 1    &       --       &       --       &       48.9607 &       0.0000  &       48.9607 &       48.9812 &       0.027   &       0.027   \\
1, 86   $\xleftarrow{}$ 2, 1    &       --       &       --       &       48.9607 &       0.8673  &       48.0934 &       48.1134 &       0.061   &       0.061   \\
2, 118  $\xleftarrow{}$ 1, 1    &       --       &       --       &       49.5105 &       0.2891  &           49.2214 &       49.2419 &       0.052   &       0.052   \\
\hline\\[-0.35cm]
\hline\\[-0.35cm]
\end{tabular}

\begin{flushleft}
%
%$^\text{a}$~
%
\end{flushleft}
}
\end{table}

%%%%%%%%%%%%%%%%%%%%%%%%%%%%%%%%%%%%%%%%%%%%%%%%%%%%%%%%%%%%%%%%%%%%%%%%%%%%%%%%%%%%%%%%%%%%%
%
% MW
%
%%%%%%%%%%%%%%%%%%%%%%%%%%%%%%%%%%%%%%%%%%%%%%%%%%%%%%%%%%%%%%%%%%%%%%%%%%%%%%%%%%%%%%%%%%%%%
\clearpage
\section{Microwave spectrum \label{ch:mirspectrum}}
\noindent
Similarly to the far-infrared high-resolution spectroscopy experiments, comparison can be made (Table~\ref{tab:MW}) with microwave observations by Suenram $et \ al.$ \cite{SuFrLoKa94}. In the microwave spectrum, there were four $\Sigma$ and six $\Pi$ bands observed at $\sim 1 K$ rotational temperature in the supersonic expansion.  
There is an excellent agreement with the experimental and also with the computed transitions by Wang and Carrington \cite{WaCa21}.

\begin{table}%[htbp]
\caption{%
Rotation-vibration-tunneling transitions of CH$_4$--H$_2$O observed in microwave spectroscopy experiments \cite{SuFrLoKa94}.
See also caption to Table~\ref{tab:E1E2}.
\label{tab:MW}  
}
\small{
\begin{tabular}{@{}l@{\ \ } c@{\ \ }c@{\ \ \ }  c@{\ \ } c@{\ \ \ } c@{\ \ } c@{\ \ }c@{\ \ } c@{}}
\hline\\[-0.35cm]
\hline\\[-0.35cm]
$(J,n)'$ $\xleftarrow{}$  $(J,n)$ &
$\Delta\tilde\nu_\text{obs}$~[MHz] &  $\Delta\tilde\nu_\text{obs}$  &
$E'$ & $E$ 
& $\Delta\tilde\nu$ & $\Delta\tilde\nu$ &
$S_0$ & 
$S_0$ \\
\\[-0.35cm]
&
\cite{DoSa94} & &
[this w]& [this w]& [this w] & \cite{WaCa21} &
[this w] & 
\cite{WaCa21} \\
\\[-0.35cm]
\hline\\[-0.35cm]
& \multicolumn{8}{l}{$A^+$, $\Sigma $ band:} \\
1, 1    $\xleftarrow{}$ 0, 1    &       8692.96 &       0.2900  &       0.2891  &       0.0000  &       0.2891  &       0.2893  &       0.040   &       0.040   \\
2, 1    $\xleftarrow{}$ 1, 1    &       17383.1 &       0.5798  &       0.8673  &       0.2891  &       0.5781  &       0.5785  &       0.079   &       0.079   \\
\\
& \multicolumn{8}{l}{$F^-$, $\Sigma $ band:} \\
1, 12--14        $\xleftarrow{}$ 0, 6--8  &       8504.65 &       0.2837  &       11.5324 &       11.2507 &       0.2817  &       0.2819  &       0.029   &       0.029   \\
2, 12--14        $\xleftarrow{}$ 1, 12--14        &       17007.9 &       0.567     &       12.0959 &       11.5324 &       0.5634  &       0.5639  &       0.059   &       0.059   \\
\\
& \multicolumn{8}{l}{$F^+$, $\Sigma $ band:} \\
1, 2--4  $\xleftarrow{}$ 0, 2--4  &       8476.93 &       0.2828  &       5.0455  &       4.7642  &       0.2814  &       0.2816  &       0.055   &       0.055   \\
2, 2--4  $\xleftarrow{}$ 1, 2--4  &       16953.0 &       0.5655  &       5.6083  &       5.0455  &       0.5628  &       0.5632  &       0.109   &       0.109   \\
\\
& \multicolumn{8}{l}{$A^-$, $\Sigma $ band:} \\
1, 5    $\xleftarrow{}$ 0, 5    &       8690.39 &       0.2899  &       7.2805  &       6.9919  &       0.2886  &       0.2888  &       0.024   &       0.024   \\
2, 5    $\xleftarrow{}$ 1, 5    &       17378.1 &       0.5797  &       7.8576  &       7.2805  &       0.5771  &       0.5774  &       0.048   &       0.048   \\
\\
& \multicolumn{8}{l}{$E^+$, $\Pi $ band:} \\
2, 15--16        $\xleftarrow{}$ 1, 15--16        &       17120.06        &       0.5711  &       13.6782 &       13.1087 &       0.5695  &       0.5700  &       0.099   &       0.099   \\
\\
& \multicolumn{8}{l}{$E^-$, $\Pi $ band:} \\
2, 23--24        $\xleftarrow{}$ 1, 23--24        &       17105.28        &       0.5706  &       19.9581 &       19.3909 &       0.5671  &       0.5676  &       0.052   &       0.052   \\
\\
& \multicolumn{8}{l}{$F^+$, $\Pi $ band:} \\
2, 6--8  $\xleftarrow{}$ 1, 6--8  &       17421.68        &       0.5811  &       8.4916  &       7.9119  &       0.5797  &       0.5800  &       0.072   &       0.071   \\
2, 9--11 $\xleftarrow{}$ 1, 9--11 &       17607.7 &       0.5873  &       8.5026  &       7.9158  &       0.5868  &       0.5873  &       0.072   &       0.071   \\
\\
& \multicolumn{8}{l}{$F^-$, $\Pi $ band:} \\
2, 17--19        $\xleftarrow{}$ 1, 17--19        &       17396.81        &       0.5803  &       15.1903 &       14.6124 &       0.5779  &       0.5780  &       0.040   &       0.040   \\
2, 20--22        $\xleftarrow{}$ 1, 20--22        &       17516.8 &       0.5843  &       15.1975 &       14.6152 &       0.5823  &       0.5828  &       0.040   &       0.040   \\
\hline\\[-0.35cm]
\hline\\[-0.35cm]
\end{tabular}

\begin{flushleft}
%
%$^\text{a}$~
%
\end{flushleft}
}
\end{table}

%%%%%%%%%%%%%%%%%%%%%%%%%%%%%%%%%%%%%%%%%%%%%%%%%%%%%%%%%%%%%%%%%%%%%%%%%%%%%%%%%%%%%%%%%%%%%
%
% Summary and conclusion
%
%%%%%%%%%%%%%%%%%%%%%%%%%%%%%%%%%%%%%%%%%%%%%%%%%%%%%%%%%%%%%%%%%%%%%%%%%%%%%%%%%%%%%%%%%%%%%
\clearpage
\section{Summary, conclusion, and outlook}
\noindent
The black-box-type rovibrational method implemented in the GENIUSH program package has been extensively tested with respect to the sophisticated dimer Hamiltonian approach that has been tailored for describing the intermolecular dynamics of floppy dimers by Wang and Carrington \cite{WaCa21}. 

GENIUSH uses a numerical kinetic energy operator approach, user-defined coordinates and body-fixed frame, and a direct product basis and grid.
The dimer approach of Wang and Carrington uses an analytic kinetic energy operator, (non-direct product) coupled basis functions including Wigner's $D$ functions, and analytic kinetic energy operator matrix elements. 

We show for the example of the rovibrational states and transitions of the methane-water dimer that the performance of the black-box-type approach is on the same order of magnitude as that of the tailor-made approach, the latter being more efficient. 
In our direct-product approach, it is important to use the sine-cot-DVR developed by Schiffel and Manthe \cite{ScMa10} for the singular `bending' coordinates.
Then, we can converge the energy levels on the order of $10^{-3}$~\cm\ with a basis set only twice as large as the non-product dimer basis of Wang and Carrington \cite{WaCa21}. (While our grid size is the same as our basis size, their grid size is unknown to us---most likely larger than their basis set size---that is the bottleneck of matrix-vector computations.) The well-converged data allowed us to spot that they should have used a slightly larger basis (and/or grid) for the intermolecular stretching degree of freedom, because  they have an error of 0.02~\cm\ in the intermolecular stretching fundamental vibration energy. 

To compute rovibrational transitions for $J=0,1,$ and 2 rotational quantum numbers, we use a basis set that is ca.~4.5 times as large as the non-product dimer basis of Wang and Carrington, and we evaluate rovibrational transition energies and line strengths. Our transitions energies are in an agreement of $10^{-3}$~\cm\, often just a few $10^{-4}$~\cm of the energies reported by Wang and Carrington. 
These deviations are far beyond the theoretical uncertainty of the current rovibrational theoretical framework based on the non-relativistic and the Born--Oppenheimer approximations.

Regarding, further progress that we would be happy to witness in the forthcoming years,  is related to  computing extensive line lists for floppy systems. An extensive line list---that can be useful for simulating molecular interactions \cite{OwYa18} or providing datasets \cite{exomol} to other disciplines---includes rotational excitations up to $J=20$ and beyond. If we have to solve the the full rovibrational problem (from scratch) for every $J$, then a factor of 2-3-5 in the vibrational basis can be an important advantage (of the tailor-made method). A more efficient approach would be, however, if we had an method for finding a good body-fixed frame that allows one to use efficiently the knowledge of the solution of the $J=0$ problem, the vibrational eigenstates in the energy range relevant for the dynamics. It may be the numerical KEO approach (also implemented in GENIUSH) that may allow the optimization of the body-fixed frame even for floppy systems \cite{AvMSDMe21}.

\vspace{1cm}
All data that support findings of this study are included in the paper and in the Supplementary
Material.

%%%%%%%%%%%%%%%%%%%%%%%%%%%%%%%%%%%%%%%%%%%%%%%%%%%%%%%%%%%%%%%%%%%%%%%%%%%%%%%%%%%%%%%%%%%%%
%
% Acknowledgment
%
%%%%%%%%%%%%%%%%%%%%%%%%%%%%%%%%%%%%%%%%%%%%%%%%%%%%%%%%%%%%%%%%%%%%%%%%%%%%%%%%%%%%%%%%%%%%%
% \clearpage
\section{Acknowledgment}
\noindent %
We thank the financial support of the Swiss National Science Foundation 
(PROMYS Grant, No.~IZ11Z0\_166525).
We are grateful to Xiao-Gang Wang and Tucker Carrington for the thorough documentation 
of their work in Ref.~\cite{WaCa21} and for paying attention to using exactly the same constant parameters as in Ref.~\cite{SaCsAlWaMa16} that made it possible for us to extensively test and further develop our methodologies.
We also thank them for sharing their sine-cot-DVR implementation during our earlier work \cite{AvMa19} that allowed us to test our computations also used in the present work.

%%%%%%%%%%%%%%%%%%%%%%%%%%%%%%%%%%%%%%%%%%%%%%%%%%%%%%%%%%%%%%%%%%%%%%%%%%%%%%%%%%%%%%%%%%%%%
%
% Bibliography
%
%%%%%%%%%%%%%%%%%%%%%%%%%%%%%%%%%%%%%%%%%%%%%%%%%%%%%%%%%%%%%%%%%%%%%%%%%%%%%%%%%%%%%%%%%%%%%
% \clearpage
% \bibliography{mybib}

\begin{thebibliography}{47}%
\makeatletter
\providecommand \@ifxundefined [1]{%
 \@ifx{#1\undefined}
}%
\providecommand \@ifnum [1]{%
 \ifnum #1\expandafter \@firstoftwo
 \else \expandafter \@secondoftwo
 \fi
}%
\providecommand \@ifx [1]{%
 \ifx #1\expandafter \@firstoftwo
 \else \expandafter \@secondoftwo
 \fi
}%
\providecommand \natexlab [1]{#1}%
\providecommand \enquote  [1]{``#1''}%
\providecommand \bibnamefont  [1]{#1}%
\providecommand \bibfnamefont [1]{#1}%
\providecommand \citenamefont [1]{#1}%
\providecommand \href@noop [0]{\@secondoftwo}%
\providecommand \href [0]{\begingroup \@sanitize@url \@href}%
\providecommand \@href[1]{\@@startlink{#1}\@@href}%
\providecommand \@@href[1]{\endgroup#1\@@endlink}%
\providecommand \@sanitize@url [0]{\catcode `\\12\catcode `\$12\catcode
  `\&12\catcode `\#12\catcode `\^12\catcode `\_12\catcode `\%12\relax}%
\providecommand \@@startlink[1]{}%
\providecommand \@@endlink[0]{}%
\providecommand \url  [0]{\begingroup\@sanitize@url \@url }%
\providecommand \@url [1]{\endgroup\@href {#1}{\urlprefix }}%
\providecommand \urlprefix  [0]{URL }%
\providecommand \Eprint [0]{\href }%
\providecommand \doibase [0]{https://doi.org/}%
\providecommand \selectlanguage [0]{\@gobble}%
\providecommand \bibinfo  [0]{\@secondoftwo}%
\providecommand \bibfield  [0]{\@secondoftwo}%
\providecommand \translation [1]{[#1]}%
\providecommand \BibitemOpen [0]{}%
\providecommand \bibitemStop [0]{}%
\providecommand \bibitemNoStop [0]{.\EOS\space}%
\providecommand \EOS [0]{\spacefactor3000\relax}%
\providecommand \BibitemShut  [1]{\csname bibitem#1\endcsname}%
\let\auto@bib@innerbib\@empty
%</preamble>
\bibitem [{\citenamefont {M\'atyus}\ \emph {et~al.}(2009)\citenamefont
  {M\'atyus}, \citenamefont {Czak\'o},\ and\ \citenamefont
  {Cs\'asz\'ar}}]{MaCzCs09}%
  \BibitemOpen
  \bibfield  {author} {\bibinfo {author} {\bibfnamefont {E.}~\bibnamefont
  {M\'atyus}}, \bibinfo {author} {\bibfnamefont {G.}~\bibnamefont {Czak\'o}},\
  and\ \bibinfo {author} {\bibfnamefont {A.~G.}\ \bibnamefont {Cs\'asz\'ar}},\
  }\bibfield  {title} {\bibinfo {title} {Toward black-box-type full- and
  reduced-dimensional variational (ro)vibrational computations},\ }\href
  {https://doi.org/10.1063/1.3076742} {\bibfield  {journal} {\bibinfo
  {journal} {J. Chem. Phys.}\ }\textbf {\bibinfo {volume} {130}},\ \bibinfo
  {pages} {134112} (\bibinfo {year} {2009})}\BibitemShut {NoStop}%
\bibitem [{\citenamefont {Meyer}\ and\ \citenamefont
  {G{\"u}nthard}(1969)}]{MeGu69}%
  \BibitemOpen
  \bibfield  {author} {\bibinfo {author} {\bibfnamefont {R.}~\bibnamefont
  {Meyer}}\ and\ \bibinfo {author} {\bibfnamefont {H.~H.}\ \bibnamefont
  {G{\"u}nthard}},\ }\bibfield  {title} {\bibinfo {title} {Internal rotation
  and vibration in {CH}$_2$={CC}l--{CH}$_2${D}},\ }\href
  {https://doi.org/10.1063/1.1670803} {\bibfield  {journal} {\bibinfo
  {journal} {J. Chem. Phys.}\ }\textbf {\bibinfo {volume} {50}},\ \bibinfo
  {pages} {353} (\bibinfo {year} {1969})}\BibitemShut {NoStop}%
\bibitem [{\citenamefont {Luckhaus}(2000)}]{Lu00}%
  \BibitemOpen
  \bibfield  {author} {\bibinfo {author} {\bibfnamefont {D.}~\bibnamefont
  {Luckhaus}},\ }\bibfield  {title} {\bibinfo {title} {{6D} vibrational quantum
  dynamics: Generalized coordinate discrete variable representation and
  (a)diabatic contraction},\ }\href {https://doi.org/10.1063/1.481924}
  {\bibfield  {journal} {\bibinfo  {journal} {J. Chem. Phys.}\ }\textbf
  {\bibinfo {volume} {113}},\ \bibinfo {pages} {1329} (\bibinfo {year}
  {2000})}\BibitemShut {NoStop}%
\bibitem [{\citenamefont {Luckhaus}(2003)}]{Lu03}%
  \BibitemOpen
  \bibfield  {author} {\bibinfo {author} {\bibfnamefont {D.}~\bibnamefont
  {Luckhaus}},\ }\bibfield  {title} {\bibinfo {title} {The vibrational spectrum
  of {HONO}: Fully coupled {6D} direct dynamics},\ }\href
  {https://doi.org/10.1063/1.1567713} {\bibfield  {journal} {\bibinfo
  {journal} {J. Chem. Phys.}\ }\textbf {\bibinfo {volume} {118}},\ \bibinfo
  {pages} {8797} (\bibinfo {year} {2003})}\BibitemShut {NoStop}%
\bibitem [{\citenamefont {Lauvergnat}\ and\ \citenamefont
  {Nauts}(2002)}]{LaNa02}%
  \BibitemOpen
  \bibfield  {author} {\bibinfo {author} {\bibfnamefont {D.}~\bibnamefont
  {Lauvergnat}}\ and\ \bibinfo {author} {\bibfnamefont {A.}~\bibnamefont
  {Nauts}},\ }\bibfield  {title} {\bibinfo {title} {Exact numerical computation
  of a kinetic energy operator in curvilinear coordinates},\ }\href
  {https://doi.org/10.1063/1.1469019} {\bibfield  {journal} {\bibinfo
  {journal} {J. Chem. Phys.}\ }\textbf {\bibinfo {volume} {116}},\ \bibinfo
  {pages} {8560} (\bibinfo {year} {2002})}\BibitemShut {NoStop}%
\bibitem [{\citenamefont {Yurchenko}\ \emph {et~al.}(2007)\citenamefont
  {Yurchenko}, \citenamefont {Thiel},\ and\ \citenamefont {Jensen}}]{YuThJe07}%
  \BibitemOpen
  \bibfield  {author} {\bibinfo {author} {\bibfnamefont {S.~N.}\ \bibnamefont
  {Yurchenko}}, \bibinfo {author} {\bibfnamefont {W.}~\bibnamefont {Thiel}},\
  and\ \bibinfo {author} {\bibfnamefont {P.}~\bibnamefont {Jensen}},\
  }\bibfield  {title} {\bibinfo {title} {Theoretical {ROV}ibrational {E}nergies
  ({TROVE}): {A} robust numerical approach to the calculation of rovibrational
  energies for polyatomic molecules},\ }\href
  {https://doi.org/10.1016/j.jms.2007.07.009} {\bibfield  {journal} {\bibinfo
  {journal} {J. Mol. Spectrosc.}\ }\textbf {\bibinfo {volume} {245}},\ \bibinfo
  {pages} {126} (\bibinfo {year} {2007})}\BibitemShut {NoStop}%
\bibitem [{\citenamefont {Brocks}\ \emph {et~al.}(1983)\citenamefont {Brocks},
  \citenamefont {{van der Avoird}}, \citenamefont {Sutcliffe},\ and\
  \citenamefont {Tennyson}}]{BrAvSuTe83}%
  \BibitemOpen
  \bibfield  {author} {\bibinfo {author} {\bibfnamefont {G.}~\bibnamefont
  {Brocks}}, \bibinfo {author} {\bibfnamefont {A.}~\bibnamefont {{van der
  Avoird}}}, \bibinfo {author} {\bibfnamefont {B.~T.}\ \bibnamefont
  {Sutcliffe}},\ and\ \bibinfo {author} {\bibfnamefont {J.}~\bibnamefont
  {Tennyson}},\ }\bibfield  {title} {\bibinfo {title} {Quantum dynamics of
  non-rigid systems comprising two polyatomic fragments},\ }\href
  {https://doi.org/10.1080/00268978300102831} {\bibfield  {journal} {\bibinfo
  {journal} {Mol. Phys.}\ }\textbf {\bibinfo {volume} {50}},\ \bibinfo {pages}
  {1025} (\bibinfo {year} {1983})}\BibitemShut {NoStop}%
\bibitem [{\citenamefont {Bačić}\ and\ \citenamefont {Light}(1989)}]{BaLi89}%
  \BibitemOpen
  \bibfield  {author} {\bibinfo {author} {\bibfnamefont {Z.}~\bibnamefont
  {Bačić}}\ and\ \bibinfo {author} {\bibfnamefont {J.~C.}\ \bibnamefont
  {Light}},\ }\bibfield  {title} {\bibinfo {title} {Theoretical methods for
  rovibrational states of floppy molecules},\ }\href
  {https://doi.org/10.1146/annurev.pc.40.100189.002345} {\bibfield  {journal}
  {\bibinfo  {journal} {Annu. Rev. Phys. Chem.}\ }\textbf {\bibinfo {volume}
  {40}},\ \bibinfo {pages} {469} (\bibinfo {year} {1989})}\BibitemShut
  {NoStop}%
\bibitem [{\citenamefont {Sutcliffe}\ and\ \citenamefont
  {Tennyson}(1991)}]{SuTe91}%
  \BibitemOpen
  \bibfield  {author} {\bibinfo {author} {\bibfnamefont {B.~T.}\ \bibnamefont
  {Sutcliffe}}\ and\ \bibinfo {author} {\bibfnamefont {J.}~\bibnamefont
  {Tennyson}},\ }\bibfield  {title} {\bibinfo {title} {A general treatment of
  vibration‐rotation coordinates for triatomic molecules},\ }\href
  {https://doi.org/10.1002/qua.560390208} {\bibfield  {journal} {\bibinfo
  {journal} {Int. J. Quant. Chem.}\ }\textbf {\bibinfo {volume} {39}},\
  \bibinfo {pages} {183} (\bibinfo {year} {1991})}\BibitemShut {NoStop}%
\bibitem [{\citenamefont {Bramley}\ and\ \citenamefont {{Carrington,
  Jr.}}(1994)}]{BrCa94}%
  \BibitemOpen
  \bibfield  {author} {\bibinfo {author} {\bibfnamefont {M.~J.}\ \bibnamefont
  {Bramley}}\ and\ \bibinfo {author} {\bibfnamefont {T.}~\bibnamefont
  {{Carrington, Jr.}}},\ }\bibfield  {title} {\bibinfo {title} {A general
  discrete variable method to calculate vibrational energy levels of three‐
  and four‐atom molecules},\ }\href {https://doi.org/10.1063/1.465576}
  {\bibfield  {journal} {\bibinfo  {journal} {J. Chem. Phys.}\ }\textbf
  {\bibinfo {volume} {99}},\ \bibinfo {pages} {8519} (\bibinfo {year}
  {1994})}\BibitemShut {NoStop}%
\bibitem [{\citenamefont {Leforestier}(1994)}]{Le94}%
  \BibitemOpen
  \bibfield  {author} {\bibinfo {author} {\bibfnamefont {C.}~\bibnamefont
  {Leforestier}},\ }\bibfield  {title} {\bibinfo {title} {Grid method for the
  {W}igner functions. application to the van der {W}aals system
  {A}r--{H}$_2${O}},\ }\href {https://doi.org/10.1063/1.468455} {\bibfield
  {journal} {\bibinfo  {journal} {J. Chem. Phys.}\ }\textbf {\bibinfo {volume}
  {101}},\ \bibinfo {pages} {7357} (\bibinfo {year} {1994})}\BibitemShut
  {NoStop}%
\bibitem [{\citenamefont {Althorpe}\ and\ \citenamefont
  {Clary}(1994)}]{AlCl94}%
  \BibitemOpen
  \bibfield  {author} {\bibinfo {author} {\bibfnamefont {S.~C.}\ \bibnamefont
  {Althorpe}}\ and\ \bibinfo {author} {\bibfnamefont {D.~C.}\ \bibnamefont
  {Clary}},\ }\bibfield  {title} {\bibinfo {title} {Calculation of the
  intermolecular bound states for water dimer},\ }\href
  {https://doi.org/10.1063/1.467545} {\bibfield  {journal} {\bibinfo  {journal}
  {J. Chem. Phys.}\ }\textbf {\bibinfo {volume} {101}},\ \bibinfo {pages}
  {3603} (\bibinfo {year} {1994})}\BibitemShut {NoStop}%
\bibitem [{\citenamefont {Zhang}\ \emph {et~al.}(1995)\citenamefont {Zhang},
  \citenamefont {Wu}, \citenamefont {Zhang}, \citenamefont {von Dirke},\ and\
  \citenamefont {Bačić}}]{ZhWuZhDiBa95}%
  \BibitemOpen
  \bibfield  {author} {\bibinfo {author} {\bibfnamefont {D.~H.}\ \bibnamefont
  {Zhang}}, \bibinfo {author} {\bibfnamefont {Q.}~\bibnamefont {Wu}}, \bibinfo
  {author} {\bibfnamefont {J.~Z.~H.}\ \bibnamefont {Zhang}}, \bibinfo {author}
  {\bibfnamefont {M.}~\bibnamefont {von Dirke}},\ and\ \bibinfo {author}
  {\bibfnamefont {Z.}~\bibnamefont {Bačić}},\ }\bibfield  {title} {\bibinfo
  {title} {Exact full‐dimensional bound state calculations for ({HF})$_2$,
  ({DF})$_2$, and {HFDF}},\ }\href {https://doi.org/10.1063/1.468719}
  {\bibfield  {journal} {\bibinfo  {journal} {J. Chem. Phys.}\ }\textbf
  {\bibinfo {volume} {102}},\ \bibinfo {pages} {2315} (\bibinfo {year}
  {1995})}\BibitemShut {NoStop}%
\bibitem [{\citenamefont {Mladenović}(2002)}]{Ml02}%
  \BibitemOpen
  \bibfield  {author} {\bibinfo {author} {\bibfnamefont {M.}~\bibnamefont
  {Mladenović}},\ }\bibfield  {title} {\bibinfo {title} {Discrete variable
  approaches to tetratomic molecules: {P}art {I}: {DVR}(6) and {DVR}(3)+{DGB}
  methods},\ }\href {https://doi.org/10.1016/S1386-1425(01)00669-2} {\bibfield
  {journal} {\bibinfo  {journal} {Spectrochim. Acta}\ }\textbf {\bibinfo
  {volume} {58}},\ \bibinfo {pages} {795} (\bibinfo {year} {2002})}\BibitemShut
  {NoStop}%
\bibitem [{\citenamefont {Wang}\ and\ \citenamefont {{Carrington,
  Jr.}}(2004)}]{WaCa04}%
  \BibitemOpen
  \bibfield  {author} {\bibinfo {author} {\bibfnamefont {X.-G.}\ \bibnamefont
  {Wang}}\ and\ \bibinfo {author} {\bibfnamefont {T.}~\bibnamefont
  {{Carrington, Jr.}}},\ }\bibfield  {title} {\bibinfo {title} {Contracted
  basis {L}anczos methods for computing numerically exact rovibrational levels
  of methane},\ }\href {https://doi.org/10.1063/1.1767093} {\bibfield
  {journal} {\bibinfo  {journal} {J. Chem. Phys.}\ }\textbf {\bibinfo {volume}
  {121}},\ \bibinfo {pages} {2937} (\bibinfo {year} {2004})}\BibitemShut
  {NoStop}%
\bibitem [{\citenamefont {Watson}(1968)}]{Wa68}%
  \BibitemOpen
  \bibfield  {author} {\bibinfo {author} {\bibfnamefont {J.~K.~G.}\
  \bibnamefont {Watson}},\ }\bibfield  {title} {\bibinfo {title}
  {Simplification of the molecular vibration-rotation hamiltonian},\ }\href
  {https://doi.org/10.1080/00268976800101381} {\bibfield  {journal} {\bibinfo
  {journal} {Mol. Phys.}\ }\textbf {\bibinfo {volume} {15}},\ \bibinfo {pages}
  {479} (\bibinfo {year} {1968})}\BibitemShut {NoStop}%
\bibitem [{\citenamefont {Whitehead}\ and\ \citenamefont
  {Handy}(1975)}]{WhHa75}%
  \BibitemOpen
  \bibfield  {author} {\bibinfo {author} {\bibfnamefont {R.~J.}\ \bibnamefont
  {Whitehead}}\ and\ \bibinfo {author} {\bibfnamefont {N.~C.}\ \bibnamefont
  {Handy}},\ }\bibfield  {title} {\bibinfo {title} {Variational calculation of
  vibration-rotation energy levels for triatomic molecules},\ }\href@noop {}
  {\bibfield  {journal} {\bibinfo  {journal} {J. Mol. Spectrosc.}\ }\textbf
  {\bibinfo {volume} {55}},\ \bibinfo {pages} {356} (\bibinfo {year}
  {1975})}\BibitemShut {NoStop}%
\bibitem [{\citenamefont {Bowman}\ \emph {et~al.}(2003)\citenamefont {Bowman},
  \citenamefont {Carter},\ and\ \citenamefont {Huang}}]{MM2}%
  \BibitemOpen
  \bibfield  {author} {\bibinfo {author} {\bibfnamefont {J.~M.}\ \bibnamefont
  {Bowman}}, \bibinfo {author} {\bibfnamefont {S.}~\bibnamefont {Carter}},\
  and\ \bibinfo {author} {\bibfnamefont {X.}~\bibnamefont {Huang}},\ }\bibfield
   {title} {\bibinfo {title} {Multimode: A code to calculate rovibrational
  energies of polyatomic molecules},\ }\href@noop {} {\bibfield  {journal}
  {\bibinfo  {journal} {Int. Rev. Phys. Chem.}\ }\textbf {\bibinfo {volume}
  {22}},\ \bibinfo {pages} {533} (\bibinfo {year} {2003})}\BibitemShut
  {NoStop}%
\bibitem [{\citenamefont {Avila}\ and\ \citenamefont
  {T.~Carrington}(2009)}]{tc-gab1}%
  \BibitemOpen
  \bibfield  {author} {\bibinfo {author} {\bibfnamefont {G.}~\bibnamefont
  {Avila}}\ and\ \bibinfo {author} {\bibfnamefont {J.}~\bibnamefont
  {T.~Carrington}},\ }\bibfield  {title} {\bibinfo {title} {Nonproduct
  quadrature grids for solving the vibrational {S}chrödinger equation},\
  }\href@noop {} {\bibfield  {journal} {\bibinfo  {journal} {J. Chem. Phys.}\
  }\textbf {\bibinfo {volume} {131}},\ \bibinfo {pages} {174103} (\bibinfo
  {year} {2009})}\BibitemShut {NoStop}%
\bibitem [{\citenamefont {Avila}\ and\ \citenamefont
  {T.~Carrington}(2011)}]{tc-gab2}%
  \BibitemOpen
  \bibfield  {author} {\bibinfo {author} {\bibfnamefont {G.}~\bibnamefont
  {Avila}}\ and\ \bibinfo {author} {\bibfnamefont {J.}~\bibnamefont
  {T.~Carrington}},\ }\bibfield  {title} {\bibinfo {title} {Using nonproduct
  quadrature grids to solve the vibrational {S}chrödinger equation in {12D}},\
  }\href@noop {} {\bibfield  {journal} {\bibinfo  {journal} {J. Chem. Phys.}\
  }\textbf {\bibinfo {volume} {134}},\ \bibinfo {pages} {054126} (\bibinfo
  {year} {2011})}\BibitemShut {NoStop}%
\bibitem [{\citenamefont {Avila}\ and\ \citenamefont
  {Mátyus}(2019{\natexlab{a}})}]{AvMa19}%
  \BibitemOpen
  \bibfield  {author} {\bibinfo {author} {\bibfnamefont {G.}~\bibnamefont
  {Avila}}\ and\ \bibinfo {author} {\bibfnamefont {E.}~\bibnamefont
  {Mátyus}},\ }\bibfield  {title} {\bibinfo {title} {Toward breaking the curse
  of dimensionality in (ro)vibrational computations of molecular systems with
  multiple large-amplitude motions},\ }\href
  {https://doi.org/10.1063/1.5090846} {\bibfield  {journal} {\bibinfo
  {journal} {J. Chem. Phys.}\ }\textbf {\bibinfo {volume} {150}},\ \bibinfo
  {pages} {174107} (\bibinfo {year} {2019}{\natexlab{a}})}\BibitemShut
  {NoStop}%
\bibitem [{\citenamefont {Avila}\ and\ \citenamefont
  {Mátyus}(2019{\natexlab{b}})}]{AvMa19b}%
  \BibitemOpen
  \bibfield  {author} {\bibinfo {author} {\bibfnamefont {G.}~\bibnamefont
  {Avila}}\ and\ \bibinfo {author} {\bibfnamefont {E.}~\bibnamefont
  {Mátyus}},\ }\bibfield  {title} {\bibinfo {title} {Full-dimensional (12{D})
  variational vibrational states of {CH}$_4\cdot${F}$^-$: interplay of
  anharmonicity and tunneling},\ }\href {https://doi.org/10.1063/1.5124532}
  {\bibfield  {journal} {\bibinfo  {journal} {J. Chem. Phys.}\ }\textbf
  {\bibinfo {volume} {151}},\ \bibinfo {pages} {154301} (\bibinfo {year}
  {2019}{\natexlab{b}})}\BibitemShut {NoStop}%
\bibitem [{\citenamefont {Avila}\ \emph {et~al.}(2020)\citenamefont {Avila},
  \citenamefont {Papp}, \citenamefont {Czakó},\ and\ \citenamefont
  {Mátyus}}]{AvPaCzMa20}%
  \BibitemOpen
  \bibfield  {author} {\bibinfo {author} {\bibfnamefont {G.}~\bibnamefont
  {Avila}}, \bibinfo {author} {\bibfnamefont {D.}~\bibnamefont {Papp}},
  \bibinfo {author} {\bibfnamefont {G.}~\bibnamefont {Czakó}},\ and\ \bibinfo
  {author} {\bibfnamefont {E.}~\bibnamefont {Mátyus}},\ }\bibfield  {title}
  {\bibinfo {title} {Exact quantum dynamics background of dispersion
  interactions: case study for {CH}$_4\cdot${A}r in full (12) dimensions},\
  }\href {https://doi.org/10.1039/C9CP04426D} {\bibfield  {journal} {\bibinfo
  {journal} {Phys. Chem. Chem. Phys.}\ }\textbf {\bibinfo {volume} {22}},\
  \bibinfo {pages} {2792} (\bibinfo {year} {2020})}\BibitemShut {NoStop}%
\bibitem [{\citenamefont {Wang}\ and\ \citenamefont
  {T.~Carrington}(2021)}]{WaCa21}%
  \BibitemOpen
  \bibfield  {author} {\bibinfo {author} {\bibfnamefont {X.-G.}\ \bibnamefont
  {Wang}}\ and\ \bibinfo {author} {\bibfnamefont {J.}~\bibnamefont
  {T.~Carrington}},\ }\bibfield  {title} {\bibinfo {title} {Using nondirect
  product {W}igner {$D$} basis functions and the symmetry adapted lanczos
  algorithm to compute the ro-vibrational spectrum of {CH}$_4$--{H}$_2${O}},\
  }\href {https://doi.org/10.1063/5.0044010} {\bibfield  {journal} {\bibinfo
  {journal} {J. Chem. Phys.}\ }\textbf {\bibinfo {volume} {154}},\ \bibinfo
  {pages} {124112} (\bibinfo {year} {2021})}\BibitemShut {NoStop}%
\bibitem [{\citenamefont {Sarka}\ \emph {et~al.}(2016)\citenamefont {Sarka},
  \citenamefont {Császár}, \citenamefont {Althorpe}, \citenamefont {Wales},\
  and\ \citenamefont {Mátyus}}]{SaCsAlWaMa16}%
  \BibitemOpen
  \bibfield  {author} {\bibinfo {author} {\bibfnamefont {J.}~\bibnamefont
  {Sarka}}, \bibinfo {author} {\bibfnamefont {A.~G.}\ \bibnamefont
  {Császár}}, \bibinfo {author} {\bibfnamefont {S.~C.}\ \bibnamefont
  {Althorpe}}, \bibinfo {author} {\bibfnamefont {D.~J.}\ \bibnamefont
  {Wales}},\ and\ \bibinfo {author} {\bibfnamefont {E.}~\bibnamefont
  {Mátyus}},\ }\bibfield  {title} {\bibinfo {title} {Rovibrational transitions
  of the methane-water dimer from intermolecular quantum dynamical
  computations},\ }\href@noop {} {\bibfield  {journal} {\bibinfo  {journal}
  {Phys. Chem. Chem. Phys.}\ }\textbf {\bibinfo {volume} {18}},\ \bibinfo
  {pages} {22816} (\bibinfo {year} {2016})}\BibitemShut {NoStop}%
\bibitem [{\citenamefont {Sarka}\ \emph {et~al.}(2017)\citenamefont {Sarka},
  \citenamefont {Császár},\ and\ \citenamefont {Mátyus}}]{SaCsMa17}%
  \BibitemOpen
  \bibfield  {author} {\bibinfo {author} {\bibfnamefont {J.}~\bibnamefont
  {Sarka}}, \bibinfo {author} {\bibfnamefont {A.~G.}\ \bibnamefont
  {Császár}},\ and\ \bibinfo {author} {\bibfnamefont {E.}~\bibnamefont
  {Mátyus}},\ }\bibfield  {title} {\bibinfo {title} {Rovibrational quantum
  dynamical computations for deuterated isotopologues of the methane–water
  dimer},\ }\href@noop {} {\bibfield  {journal} {\bibinfo  {journal} {Phys.
  Chem. Chem. Phys.}\ }\textbf {\bibinfo {volume} {19}},\ \bibinfo {pages}
  {15335} (\bibinfo {year} {2017})}\BibitemShut {NoStop}%
\bibitem [{\citenamefont {Metz}\ \emph {et~al.}(2019)\citenamefont {Metz},
  \citenamefont {Szalewicz}, \citenamefont {Sarka}, \citenamefont {Tóbiás},
  \citenamefont {Császár},\ and\ \citenamefont {Mátyus}}]{dimers}%
  \BibitemOpen
  \bibfield  {author} {\bibinfo {author} {\bibfnamefont {M.~P.}\ \bibnamefont
  {Metz}}, \bibinfo {author} {\bibfnamefont {K.}~\bibnamefont {Szalewicz}},
  \bibinfo {author} {\bibfnamefont {J.}~\bibnamefont {Sarka}}, \bibinfo
  {author} {\bibfnamefont {R.}~\bibnamefont {Tóbiás}}, \bibinfo {author}
  {\bibfnamefont {A.~G.}\ \bibnamefont {Császár}},\ and\ \bibinfo {author}
  {\bibfnamefont {E.}~\bibnamefont {Mátyus}},\ }\bibfield  {title} {\bibinfo
  {title} {Molecular dimers of methane clathrates: ab initio potential energy
  surfaces and variational vibrational states},\ }\href
  {https://doi.org/10.1039/C9CP00993K} {\bibfield  {journal} {\bibinfo
  {journal} {Phys. Chem. Chem. Phys.}\ }\textbf {\bibinfo {volume} {21}},\
  \bibinfo {pages} {13504} (\bibinfo {year} {2019})}\BibitemShut {NoStop}%
\bibitem [{\citenamefont {Dore}\ \emph {et~al.}(1994)\citenamefont {Dore},
  \citenamefont {Cohen}, \citenamefont {Schmuttenmaer}, \citenamefont
  {Busarow}, \citenamefont {Elrod}, \citenamefont {Loeser},\ and\ \citenamefont
  {Saykally}}]{DoSa94}%
  \BibitemOpen
  \bibfield  {author} {\bibinfo {author} {\bibfnamefont {L.}~\bibnamefont
  {Dore}}, \bibinfo {author} {\bibfnamefont {R.~C.}\ \bibnamefont {Cohen}},
  \bibinfo {author} {\bibfnamefont {C.~A.}\ \bibnamefont {Schmuttenmaer}},
  \bibinfo {author} {\bibfnamefont {K.~L.}\ \bibnamefont {Busarow}}, \bibinfo
  {author} {\bibfnamefont {M.~J.}\ \bibnamefont {Elrod}}, \bibinfo {author}
  {\bibfnamefont {J.~G.}\ \bibnamefont {Loeser}},\ and\ \bibinfo {author}
  {\bibfnamefont {R.~J.}\ \bibnamefont {Saykally}},\ }\bibfield  {title}
  {\bibinfo {title} {Far infrared vibration-rotation-tunneling spectroscopy and
  internal dynamics of methane-water: A prototypical hydrophobic system},\
  }\href {https://doi.org/10.1063/1.466569} {\bibfield  {journal} {\bibinfo
  {journal} {J. Chem. Phys.}\ }\textbf {\bibinfo {volume} {100}},\ \bibinfo
  {pages} {863} (\bibinfo {year} {1994})}\BibitemShut {NoStop}%
\bibitem [{\citenamefont {Suenram}\ \emph {et~al.}()\citenamefont {Suenram},
  \citenamefont {Fraser}, \citenamefont {Lovas},\ and\ \citenamefont
  {Kawashima}}]{SuFrLoKa94}%
  \BibitemOpen
  \bibfield  {author} {\bibinfo {author} {\bibfnamefont {R.~D.}\ \bibnamefont
  {Suenram}}, \bibinfo {author} {\bibfnamefont {G.~T.}\ \bibnamefont {Fraser}},
  \bibinfo {author} {\bibfnamefont {F.~J.}\ \bibnamefont {Lovas}},\ and\
  \bibinfo {author} {\bibfnamefont {Y.}~\bibnamefont {Kawashima}},\ }\bibfield
  {title} {\bibinfo {title} {The microwave spectrum of {CH}$_4$--{H$_2${O} },
  journal = jcp, volume = {101}, pages = {7230}, year = {1994}, doi=
  {10.1063/1.468280}},\ }\href@noop {} {\ }\BibitemShut {NoStop}%
\bibitem [{\citenamefont {Fábri}\ \emph {et~al.}(2017)\citenamefont {Fábri},
  \citenamefont {Quack},\ and\ \citenamefont {Császár}}]{FaQuCs17}%
  \BibitemOpen
  \bibfield  {author} {\bibinfo {author} {\bibfnamefont {C.}~\bibnamefont
  {Fábri}}, \bibinfo {author} {\bibfnamefont {M.}~\bibnamefont {Quack}},\ and\
  \bibinfo {author} {\bibfnamefont {A.~G.}\ \bibnamefont {Császár}},\
  }\bibfield  {title} {\bibinfo {title} {On the use of nonrigid-molecular
  symmetry in nuclear motion computations employing a discrete variable
  representation: A case study of the bending energy levels of {CH}$_5^+$},\
  }\href {https://doi.org/10.1063/1.4990297} {\bibfield  {journal} {\bibinfo
  {journal} {J. Chem. Phys.}\ }\textbf {\bibinfo {volume} {147}},\ \bibinfo
  {pages} {134101} (\bibinfo {year} {2017})}\BibitemShut {NoStop}%
\bibitem [{\citenamefont {M\'atyus}\ \emph {et~al.}(2010)\citenamefont
  {M\'atyus}, \citenamefont {F\'abri}, \citenamefont {Szidarovszky},
  \citenamefont {Czak\'o}, \citenamefont {Allen},\ and\ \citenamefont
  {Cs\'asz\'ar}}]{NMD10}%
  \BibitemOpen
  \bibfield  {author} {\bibinfo {author} {\bibfnamefont {E.}~\bibnamefont
  {M\'atyus}}, \bibinfo {author} {\bibfnamefont {C.}~\bibnamefont {F\'abri}},
  \bibinfo {author} {\bibfnamefont {T.}~\bibnamefont {Szidarovszky}}, \bibinfo
  {author} {\bibfnamefont {G.}~\bibnamefont {Czak\'o}}, \bibinfo {author}
  {\bibfnamefont {W.~D.}\ \bibnamefont {Allen}},\ and\ \bibinfo {author}
  {\bibfnamefont {A.~G.}\ \bibnamefont {Cs\'asz\'ar}},\ }\bibfield  {title}
  {\bibinfo {title} {Assigning quantum labels to variationally computed
  rotational-vibrational eigenstates of polyatomic molecules},\ }\href@noop {}
  {\bibfield  {journal} {\bibinfo  {journal} {J. Chem. Phys.}\ }\textbf
  {\bibinfo {volume} {133}},\ \bibinfo {pages} {034113} (\bibinfo {year}
  {2010})}\BibitemShut {NoStop}%
\bibitem [{\citenamefont {Avila}\ \emph {et~al.}()\citenamefont {Avila},
  \citenamefont {{Mart\'in Santa Dar\'ia}},\ and\ \citenamefont
  {M\'atyus}}]{AvMSDMe21}%
  \BibitemOpen
  \bibfield  {author} {\bibinfo {author} {\bibfnamefont {G.}~\bibnamefont
  {Avila}}, \bibinfo {author} {\bibfnamefont {A.}~\bibnamefont {{Mart\'in Santa
  Dar\'ia}}},\ and\ \bibinfo {author} {\bibfnamefont {E.}~\bibnamefont
  {M\'atyus}},\ }\href@noop {} {\bibinfo {title} {Optimization of the
  body-fixed frame to minimize rovibrational coupling, in preparation
  (2021)}}\BibitemShut {NoStop}%
\bibitem [{\citenamefont {Akin-Ojo}\ and\ \citenamefont
  {Szalewicz}(2005)}]{AOSz05}%
  \BibitemOpen
  \bibfield  {author} {\bibinfo {author} {\bibfnamefont {O.}~\bibnamefont
  {Akin-Ojo}}\ and\ \bibinfo {author} {\bibfnamefont {K.}~\bibnamefont
  {Szalewicz}},\ }\bibfield  {title} {\bibinfo {title} {Potential energy
  surface and second virial coefficient of methane-water from ab initio
  calculations},\ }\href@noop {} {\bibfield  {journal} {\bibinfo  {journal} {J.
  Chem. Phys.}\ }\textbf {\bibinfo {volume} {123}},\ \bibinfo {pages} {134311}
  (\bibinfo {year} {2005})}\BibitemShut {NoStop}%
\bibitem [{\citenamefont {F\'abri}\ \emph {et~al.}(2011)\citenamefont
  {F\'abri}, \citenamefont {M\'atyus},\ and\ \citenamefont
  {Cs\'asz\'ar}}]{FaMaCs11}%
  \BibitemOpen
  \bibfield  {author} {\bibinfo {author} {\bibfnamefont {C.}~\bibnamefont
  {F\'abri}}, \bibinfo {author} {\bibfnamefont {E.}~\bibnamefont {M\'atyus}},\
  and\ \bibinfo {author} {\bibfnamefont {A.~G.}\ \bibnamefont {Cs\'asz\'ar}},\
  }\bibfield  {title} {\bibinfo {title} {Rotating full- and reduced-dimensional
  quantum chemical models of molecules},\ }\href
  {https://doi.org/10.1063/1.3533950} {\bibfield  {journal} {\bibinfo
  {journal} {J. Chem. Phys.}\ }\textbf {\bibinfo {volume} {134}},\ \bibinfo
  {pages} {074105} (\bibinfo {year} {2011})}\BibitemShut {NoStop}%
\bibitem [{\citenamefont {Fábri}\ \emph {et~al.}(2013)\citenamefont {Fábri},
  \citenamefont {Császár},\ and\ \citenamefont {Czakó}}]{FaCsCz13}%
  \BibitemOpen
  \bibfield  {author} {\bibinfo {author} {\bibfnamefont {C.}~\bibnamefont
  {Fábri}}, \bibinfo {author} {\bibfnamefont {A.~G.}\ \bibnamefont
  {Császár}},\ and\ \bibinfo {author} {\bibfnamefont {G.}~\bibnamefont
  {Czakó}},\ }\bibfield  {title} {\bibinfo {title} {Reduced-dimensional
  quantum computations for the rotational-vibrational dynamics of f--ch4 and
  f--ch2d2},\ }\href@noop {} {\bibfield  {journal} {\bibinfo  {journal} {J.
  Phys. Chem. A}\ }\textbf {\bibinfo {volume} {117}},\ \bibinfo {pages} {6975}
  (\bibinfo {year} {2013})}\BibitemShut {NoStop}%
\bibitem [{\citenamefont {F\'abri}\ \emph {et~al.}(2014)\citenamefont
  {F\'abri}, \citenamefont {M\'atyus},\ and\ \citenamefont
  {Cs\'asz\'ar}}]{14FaMaCs}%
  \BibitemOpen
  \bibfield  {author} {\bibinfo {author} {\bibfnamefont {C.}~\bibnamefont
  {F\'abri}}, \bibinfo {author} {\bibfnamefont {E.}~\bibnamefont {M\'atyus}},\
  and\ \bibinfo {author} {\bibfnamefont {A.~G.}\ \bibnamefont {Cs\'asz\'ar}},\
  }\href@noop {} {\bibfield  {journal} {\bibinfo  {journal} {Spectrochim.
  Acta}\ }\textbf {\bibinfo {volume} {119}},\ \bibinfo {pages} {84} (\bibinfo
  {year} {2014})}\BibitemShut {NoStop}%
\bibitem [{\citenamefont {Papp}\ \emph {et~al.}(2017)\citenamefont {Papp},
  \citenamefont {Sarka}, \citenamefont {Szidarovszky}, \citenamefont
  {Császár}, \citenamefont {Mátyus}, \citenamefont {Hochlaf},\ and\
  \citenamefont {Stoecklin}}]{ArNOp}%
  \BibitemOpen
  \bibfield  {author} {\bibinfo {author} {\bibfnamefont {D.}~\bibnamefont
  {Papp}}, \bibinfo {author} {\bibfnamefont {J.}~\bibnamefont {Sarka}},
  \bibinfo {author} {\bibfnamefont {T.}~\bibnamefont {Szidarovszky}}, \bibinfo
  {author} {\bibfnamefont {A.~G.}\ \bibnamefont {Császár}}, \bibinfo {author}
  {\bibfnamefont {E.}~\bibnamefont {Mátyus}}, \bibinfo {author} {\bibfnamefont
  {M.}~\bibnamefont {Hochlaf}},\ and\ \bibinfo {author} {\bibfnamefont
  {T.}~\bibnamefont {Stoecklin}},\ }\bibfield  {title} {\bibinfo {title}
  {Complex rovibrational dynamics of the ar$\cdot$no$^+$ complex},\ }\href@noop
  {} {\bibfield  {journal} {\bibinfo  {journal} {Phys. Chem. Chem. Phys.}\
  }\textbf {\bibinfo {volume} {19}},\ \bibinfo {pages} {8152} (\bibinfo {year}
  {2017})}\BibitemShut {NoStop}%
\bibitem [{\citenamefont {Sarka}\ and\ \citenamefont
  {Császár}(2016)}]{SaCs16}%
  \BibitemOpen
  \bibfield  {author} {\bibinfo {author} {\bibfnamefont {J.}~\bibnamefont
  {Sarka}}\ and\ \bibinfo {author} {\bibfnamefont {A.~G.}\ \bibnamefont
  {Császár}},\ }\bibfield  {title} {\bibinfo {title} {Interpretation of the
  vibrational energy level structure of the astructural molecular ion ${H}_5^+$
  and all of its deuterated isotopomers},\ }\href
  {https://doi.org/10.1063/1.4946808} {\bibfield  {journal} {\bibinfo
  {journal} {J. Chem. Phys.}\ }\textbf {\bibinfo {volume} {144}},\ \bibinfo
  {pages} {154309} (\bibinfo {year} {2016})}\BibitemShut {NoStop}%
\bibitem [{\citenamefont {Ferenc}\ and\ \citenamefont
  {Mátyus}(2019)}]{FeMa19}%
  \BibitemOpen
  \bibfield  {author} {\bibinfo {author} {\bibfnamefont {D.}~\bibnamefont
  {Ferenc}}\ and\ \bibinfo {author} {\bibfnamefont {E.}~\bibnamefont
  {Mátyus}},\ }\bibfield  {title} {\bibinfo {title} {Bound and unbound
  rovibrational states of the methane-argon dimer},\ }\href
  {https://doi.org/10.1080/00268976.2018.1547430} {\ \textbf {\bibinfo {volume}
  {117}},\ \bibinfo {pages} {1694} (\bibinfo {year} {2019})}\BibitemShut
  {NoStop}%
\bibitem [{\citenamefont {A.~{Mart\'in Santa Dar\'ia}}\ and\ \citenamefont
  {Mátyus}(2021)}]{fad21}%
  \BibitemOpen
  \bibfield  {author} {\bibinfo {author} {\bibfnamefont {G.~A.}\ \bibnamefont
  {A.~{Mart\'in Santa Dar\'ia}}}\ and\ \bibinfo {author} {\bibfnamefont
  {E.}~\bibnamefont {Mátyus}},\ }\bibfield  {title} {\bibinfo {title}
  {Fingerprint region of the formic acid dimer: variational vibrational
  computations in curvilinear coordinates},\ }\href
  {https://doi.org/10.1039/D0CP06289H} {\bibfield  {journal} {\bibinfo
  {journal} {Phys. Chem. Chem. Phys.}\ }\textbf {\bibinfo {volume} {23}},\
  \bibinfo {pages} {6526} (\bibinfo {year} {2021})}\BibitemShut {NoStop}%
\bibitem [{\citenamefont {Meyer}(1979)}]{MeJMS79}%
  \BibitemOpen
  \bibfield  {author} {\bibinfo {author} {\bibfnamefont {R.}~\bibnamefont
  {Meyer}},\ }\bibfield  {title} {\bibinfo {title} {Flexible models for
  intramolecular motion, a versatile treatment and its application to
  glyoxal},\ }\href {https://doi.org/10.1016/0022-2852(79)90230-3} {\bibfield
  {journal} {\bibinfo  {journal} {J. Mol. Spectrosc.}\ }\textbf {\bibinfo
  {volume} {76}},\ \bibinfo {pages} {266} (\bibinfo {year} {1979})}\BibitemShut
  {NoStop}%
\bibitem [{\citenamefont {Light}\ and\ \citenamefont
  {Carrington~Jr}(2000)}]{00LiCa}%
  \BibitemOpen
  \bibfield  {author} {\bibinfo {author} {\bibfnamefont {J.~C.}\ \bibnamefont
  {Light}}\ and\ \bibinfo {author} {\bibfnamefont {T.}~\bibnamefont
  {Carrington~Jr}},\ }\bibfield  {title} {\bibinfo {title} {Discrete variable
  representations and their utilization},\ }\href
  {https://doi.org/10.1002/9780470141731.ch4} {\bibfield  {journal} {\bibinfo
  {journal} {Adv. Chem. Phys.}\ }\textbf {\bibinfo {volume} {114}},\ \bibinfo
  {pages} {263} (\bibinfo {year} {2000})}\BibitemShut {NoStop}%
\bibitem [{\citenamefont {Schiffel}\ and\ \citenamefont
  {Manthe}(2010)}]{ScMa10}%
  \BibitemOpen
  \bibfield  {author} {\bibinfo {author} {\bibfnamefont {G.}~\bibnamefont
  {Schiffel}}\ and\ \bibinfo {author} {\bibfnamefont {U.}~\bibnamefont
  {Manthe}},\ }\bibfield  {title} {\bibinfo {title} {On direct product based
  discrete variable representations for angular coordinates and the treatment
  of singular terms in the kinetic energy operator},\ }\href
  {https://doi.org/10.1016/j.chemphys.2010.07.006} {\bibfield  {journal}
  {\bibinfo  {journal} {Chem. Phys.}\ }\textbf {\bibinfo {volume} {374}},\
  \bibinfo {pages} {118} (\bibinfo {year} {2010})}\BibitemShut {NoStop}%
\bibitem [{\citenamefont {{Wolfram Research, Inc.}}()}]{WolframMath}%
  \BibitemOpen
  \bibfield  {author} {\bibinfo {author} {\bibnamefont {{Wolfram Research,
  Inc.}}},\ }\href@noop {} {\bibinfo {title} {{M}athematica, {V}ersion 12.1}},\
  \bibinfo {note} {{C}hampaign, {IL}, 2020}\BibitemShut {NoStop}%
\bibitem [{\citenamefont {Yurchenko}\ \emph {et~al.}(2009)\citenamefont
  {Yurchenko}, \citenamefont {Barber}, \citenamefont {Yachmenev}, \citenamefont
  {Thiel}, \citenamefont {Jensen},\ and\ \citenamefont {Tennyson}}]{YuBaYa09}%
  \BibitemOpen
  \bibfield  {author} {\bibinfo {author} {\bibfnamefont {S.~N.}\ \bibnamefont
  {Yurchenko}}, \bibinfo {author} {\bibfnamefont {R.~J.}\ \bibnamefont
  {Barber}}, \bibinfo {author} {\bibfnamefont {A.}~\bibnamefont {Yachmenev}},
  \bibinfo {author} {\bibfnamefont {W.}~\bibnamefont {Thiel}}, \bibinfo
  {author} {\bibfnamefont {P.}~\bibnamefont {Jensen}},\ and\ \bibinfo {author}
  {\bibfnamefont {J.}~\bibnamefont {Tennyson}},\ }\bibfield  {title} {\bibinfo
  {title} {A variationally computed {$T=300$} {K} line list for {NH}$_3$},\
  }\href {https://doi.org/10.1021/jp9029425} {\bibfield  {journal} {\bibinfo
  {journal} {J. Phys. Chem. A}\ }\textbf {\bibinfo {volume} {113}},\ \bibinfo
  {pages} {11845} (\bibinfo {year} {2009})}\BibitemShut {NoStop}%
\bibitem [{\citenamefont {Owens}\ and\ \citenamefont
  {Yachmenev}(2018)}]{OwYa18}%
  \BibitemOpen
  \bibfield  {author} {\bibinfo {author} {\bibfnamefont {A.}~\bibnamefont
  {Owens}}\ and\ \bibinfo {author} {\bibfnamefont {A.}~\bibnamefont
  {Yachmenev}},\ }\bibfield  {title} {\bibinfo {title} {Richmol: A general
  variational approach for rovibrational molecular dynamics in external
  electric fields},\ }\href {https://doi.org/10.1063/1.5023874} {\bibfield
  {journal} {\bibinfo  {journal} {J. Chem. Phys.}\ }\textbf {\bibinfo {volume}
  {148}},\ \bibinfo {pages} {124102} (\bibinfo {year} {2018})}\BibitemShut
  {NoStop}%
\bibitem [{\citenamefont {Tennyson}\ \emph {et~al.}(2016)\citenamefont
  {Tennyson}, \citenamefont {Yurchenko}, \citenamefont {Al-Refaie},
  \citenamefont {Barton}, \citenamefont {Chubb}, \citenamefont {Coles},
  \citenamefont {Diamantopoulou}, \citenamefont {Gorman}, \citenamefont {Hill},
  \citenamefont {Lam}, \citenamefont {Lodi}, \citenamefont {McKemmish},
  \citenamefont {Na}, \citenamefont {Owens}, \citenamefont {Polyansky},
  \citenamefont {Rivlin}, \citenamefont {Sousa-Silva}, \citenamefont
  {Underwood}, \citenamefont {Yachmenev},\ and\ \citenamefont {Zak}}]{exomol}%
  \BibitemOpen
  \bibfield  {author} {\bibinfo {author} {\bibfnamefont {J.}~\bibnamefont
  {Tennyson}}, \bibinfo {author} {\bibfnamefont {S.~N.}\ \bibnamefont
  {Yurchenko}}, \bibinfo {author} {\bibfnamefont {A.~F.}\ \bibnamefont
  {Al-Refaie}}, \bibinfo {author} {\bibfnamefont {E.~J.}\ \bibnamefont
  {Barton}}, \bibinfo {author} {\bibfnamefont {K.~L.}\ \bibnamefont {Chubb}},
  \bibinfo {author} {\bibfnamefont {P.~A.}\ \bibnamefont {Coles}}, \bibinfo
  {author} {\bibfnamefont {S.}~\bibnamefont {Diamantopoulou}}, \bibinfo
  {author} {\bibfnamefont {M.~N.}\ \bibnamefont {Gorman}}, \bibinfo {author}
  {\bibfnamefont {C.}~\bibnamefont {Hill}}, \bibinfo {author} {\bibfnamefont
  {A.~Z.}\ \bibnamefont {Lam}}, \bibinfo {author} {\bibfnamefont
  {L.}~\bibnamefont {Lodi}}, \bibinfo {author} {\bibfnamefont {L.~K.}\
  \bibnamefont {McKemmish}}, \bibinfo {author} {\bibfnamefont {Y.}~\bibnamefont
  {Na}}, \bibinfo {author} {\bibfnamefont {A.}~\bibnamefont {Owens}}, \bibinfo
  {author} {\bibfnamefont {O.~L.}\ \bibnamefont {Polyansky}}, \bibinfo {author}
  {\bibfnamefont {T.}~\bibnamefont {Rivlin}}, \bibinfo {author} {\bibfnamefont
  {C.}~\bibnamefont {Sousa-Silva}}, \bibinfo {author} {\bibfnamefont {D.~S.}\
  \bibnamefont {Underwood}}, \bibinfo {author} {\bibfnamefont {A.}~\bibnamefont
  {Yachmenev}},\ and\ \bibinfo {author} {\bibfnamefont {E.}~\bibnamefont
  {Zak}},\ }\bibfield  {title} {\bibinfo {title} {The exomol database:
  Molecular line lists for exoplanet and other hot atmospheres},\ }\href
  {https://doi.org/https://doi.org/10.1016/j.jms.2016.05.002} {\bibfield
  {journal} {\bibinfo  {journal} {Journal of Molecular Spectroscopy}\ }\textbf
  {\bibinfo {volume} {327}},\ \bibinfo {pages} {73} (\bibinfo {year} {2016})},\
  \bibinfo {note} {new Visions of Spectroscopic Databases, Volume
  II}\BibitemShut {NoStop}%
\end{thebibliography}

\end{document}